# Gauge invariant three boson vertices and their Ward identities in the Standard Model

Joannis Papavassiliou and Kostas Philippides

*New York University, Department of Physics*
*Andre and Bella Meyer Hall of Physics, 4 Washington Place,*
*New York, NY 10003*

## ABSTRACT

In the context of the Standard Model we extend the S–matrix pinch technique for non–conserved currents to the case of three boson vertices. We outline in detail how effective gauge invariant three boson vertices can be constructed, with all three incoming momenta *off–shell*. Explicit closed expressions for the vertices $\gamma W^- W^+$, $ZW^- W^+$, and $\chi W^- W^+$ are reported. The three boson vertices so constructed satisfy naive QED–like Ward identities which relate them to the gauge invariant gauge boson self–energies previously constructed by the same method. The derivation of the aforementioned Ward identities relies on the sole requirement of complete gauge invariance of the S–matrix element considered; in particular, no knowledge of the explicit closed form of the three boson vertices involved is necessary. The validity of one of these Ward identities is demonstrated explicitly, through a detailed diagrammatic one-loop analysis, in the context of three different gauges.

# 1 Introduction

The pinch technique (PT) is an algorithm that allows the construction of modified gauge independent (g.i.) off-shell $n$-point functions, through the order by order rearrangement of Feynman graphs contributing to a certain physical and therefore ostensibly g.i. amplitude, such as an S-matrix element or a Wilson loop [1]. The PT was originally introduced in an attempt to gain insight from perturbation theory on issues encountered in developing a consistent truncation scheme for the Schwinger-Dyson (SD) equations governing the non-perturbative QCD dynamics [2]. Specifically, one wishes to construct a SD series which is manifestly g.i. already in its one-dressed loop truncated version. This is a non-trivial task, since the mechanism of gauge cancellations is very subtle, and involves in general a delicate conspiracy of terms coming from all orders.

The systematic derivation of such a SD series for QCD has been the focal point of extensive research [3] [4]. Of particular interest in this context is the study of the three gluon vertex $\hat{\Gamma}_3$ [5], and the four gluon vertex $\hat{\Gamma}_4$ [6]. In particular, as explained first in [4] and later in [6], one attempts to construct an effective potential $\Omega$ [7] for quarkless QCD, which, in ghost-free gauges, is a functional of only three basic quantities: the gluon self-energy ($\hat{d}$), the three gluon vertex ($\hat{\Gamma}_3$), and the four gluon vertex ($\hat{\Gamma}_4$), e.g. $\Omega(\hat{d}, \hat{\Gamma}_3, \hat{\Gamma}_4)$. One then requires that $\Omega$ is manifestly gauge-independent for *off-shell* $\hat{d}$, $\hat{\Gamma}_3$, and $\hat{\Gamma}_4$, e.g. when they do not necessarily satisfy their respective SD equations. This requirement can be enforced if $\hat{d}$, $\hat{\Gamma}_3$, and $\hat{\Gamma}_4$ are individually gauge-independent, and, at the same time, the renormalized self energy $\hat{\Pi}_{\mu\nu}$ is transverse, e.g.

$$q^\mu \hat{\Pi}_{\mu\nu} = 0 \qquad (1.1)$$

order by order in the dressed loop expansion [8]. The one-loop dressed expression for $\hat{\Pi}_{\mu\nu}$ is schematically shown in Fig.1; we see that already at this level the fully dressed vertices $\hat{\Gamma}_3$ and $\hat{\Gamma}_4$ make their appearance. It turns out that Eq.(1.1) can be satisfied as long as $\hat{d}$, $\hat{\Gamma}_3$, and $\hat{\Gamma}_4$ satisfy the following Ward identities (WI):

$$q_1^\mu \hat{\Gamma}_{\mu\nu\alpha}(q_1, q_2, q_3) = T_{\nu\alpha}(q_2)\hat{d}^{-1}(q_2) - T_{\nu\alpha}(q_3)\hat{d}^{-1}(q_3) \qquad (1.2)$$

$$q_1^\mu \hat{\Gamma}_{\mu\nu\alpha\beta}^{abcd} = f_{abp}\hat{\Gamma}_{\nu\alpha\beta}^{cdp}(q_1+q_2, q_3, q_4) + c.p. \,, \qquad (1.3)$$

$$\qquad (1.4)$$

where $\hat{d}^{-1}(q) = q^2 - \hat{\Pi}(q)$, $T_{\mu\nu}(q) = g_{\mu\nu} - q_\mu q_\nu/q^2$ is the usual transverse projection operator, $f^{abc}$ the structure constants of the gauge group, and the abbreviation c.p. in the r.h.s. of Eq.(1.4) stands for "cyclic permutations" [9].

Although this program has been layed out conceptually, its practical implementation is as yet incomplete. One thing is certain however: if Green's functions with the properties described above can arise out of a self-consistent treatment of QCD, one should be able to construct Green's functions with the same properties at the level of ordinary perturbation theory after appropriate rearrangement of Feynman graphs. The PT accomplishes this task by providing the systematic algorithm needed to recover the desired Green's functions order by order in perturbation theory. So, g.i. three and four- gluon vertices have already been constructed via the PT at one-loop, and they satisfy the Ward identities of Eq.(1.4) [10].

A program similar to that outlined above for QCD has also been proposed for the case of non-Abelian gauge theories with either elementary Higgs particles or with dynamical symmetry breaking [11]. In an attempt to study the general structure of the g.i. Green's functions involved, the PT was extended to the case of theories with tree-level symmetry breaking. The technical modifications necessary to accomplish such a task have been presented in [12], in the context of an $SU(2)$ toy model. The upshot of that analysis was that the PT, when properly applied, gives rise to g.i. two- and three point functions, which satisfy the same WI as in the symmetric (unbroken) case, provided one includes appropriate longitudinal Goldstone boson Green's functions. So, for example Eq.(1.1) becomes

$$q^\mu \hat{\Pi}_{\mu\nu} + M\hat{\Pi}_\nu = 0 \tag{1.5}$$

where $\hat{\Pi}_\nu$ is the g.i one-loop mixed self–energy between the (massive) gauge boson and the corresponding Goldstone boson. Clearly, Eq.(1.1) may be enforced, if we redefine the gauge boson self-energy to be $\hat{\Pi}_{\mu\nu}^{tr} = \hat{\Pi}_{\mu\nu} + \frac{Mq_\mu}{q^2}\hat{\Pi}_\nu$ with similar redefinitions for other $n$-point functions. Subsequently, the PT was extended to the full SM [13], and several interesting applications were proposed [14], [15], [16], [17].

Even though formal considerations similar to those of the QCD case would provide sufficient grounds for a detailed study of g.i. three and four gauge boson vertices in the context of the SM, such a study was precipitated by phenomenological issues. In particular, the possibility of directly probing non-Abelian vertices in the upcoming LEP2 experiments, through the process $e^+e^- \to W^+W^-$, has led to extensive studies of anomalous gauge boson couplings, induced either by extensions of the SM, or by one-loop corrections within the SM [18], [19], [20]. In computing the latter, issues of gauge invariance become very important. So, form factors of the $W$ boson, such as the magnetic dipole and electric quadrupole moments, turn out to be gauge dependent when extracted from the conventional off-shell $\gamma WW$ and $ZWW$ vertices, calculated in the context of the $R_\xi$ gauges [22]. In addition, these quantities are infrared divergent and violate perturbative unitarity. All the above pathologies can be bypassed, as long as one instead extracts them from g.i. off-shell $\gamma WW$ and $ZWW$ vertices constructed via the PT [23].

Given the relevance of g.i three boson vertices (TBV) both from the theoretical and the phenomenological point of view, we present in this paper the general methodology for their construction, for the electro-weak sector of the SM. We focus on the vertices involving one neutral and two charged incoming particles, with all three incoming momenta off-shell. In order to construct such vertices we consider a matrix element for six-fermion elastic scattering of the form $e^-e^-\nu \to e^-e^-\nu$, where the external electrons $e$ are considered to be massive. This assumption is important, since, in addition to the g.i. vertices with three incoming gauge bosons ($\gamma W^+W^-$ and $ZW^+W^-$) it enables the construction of g.i. three boson vertices where at least one of the incoming bosons is a scalar particle (unphysical would-be Goldstone bosons and physical Higgs boson). As we will see in what follows, the latter play a crucial role in the Ward identities enforcing the gauge-invariance of the $S$-matrix In particular, in this paper we focus on the following issues.

a) We discuss the technical difficulties involved in the application of the PT when the necessary assumption is made that $m_e \neq 0$.

b) We present the most general algorithm for constructing g.i vertices involving one neutral and two charged bosons.

c) We explain how the requirement of the gauge invariance of the S-matrix gives rise to a set of WI, relating several of the g.i. vertices to each other. The derivation is general and does not require knowledge of the explicit closed form of the quantities involved. Most noticeably, the following WI,

$$q^\mu \widehat{\Gamma}_{\mu\alpha\beta}^{ZW^-W^+}(q,p_1,p_2) + iM_Z \widehat{\Gamma}_{\alpha\beta}^{\chi W^-W^+}(q,p_1,p_2) = gc \left[ \widehat{\Pi}_{\alpha\beta}^W(p_1) - \widehat{\Pi}_{\alpha\beta}^W(p_2) \right] \qquad (1.6)$$

relates the g.i vertices $\widehat{\Gamma}_{\mu\alpha\beta}^{ZW^-W^+}$ and $\widehat{\Gamma}_{\alpha\beta}^{\chi W^-W^+}$ to the g.i. $W$ self-energy $\widehat{\Pi}_{\alpha\beta}^W$. To the best of our knowledge the WI we present here has not been derived before within the PT or any other framework

There is one additional reason why the study of the g.i. vertices and WI via the PT is interesting. As it was recently realized there is a close connection between the PT and the background field method (BFM) [24]. In particular it was shown that in all cases considered so far the PT Green's functions may be obtained directly, if one computes the conventional Green's functions in the context of the BFM, using the special value $\xi_Q = 1$ of the gauge fixing parameter used to gauge-fix the quantum field [25] [26]. Since however no formal connection between the two methods has yet been established, additional cases may have to be considered, at least for those Green's functions which are of particular physical relevance. The method for constructing vertices advertised above provides the framework for such a detailed investigation.

It is important to emphasize that the closed form of the g.i. TBV obtained by the application of the S-matrix PT does *not* depend on the particular process employed. So, instead of the process $ee\nu \to ee\nu$ one could equally well extract the g.i TBV from a process of the form $bbt \to bbt$ where $t$ and $b$ are the top and bottom quarks, respectively, or even a process involving gauge bosons as external on shell particles, such as $WW\gamma \to WW\gamma$. The fact that the PT gives rise to *process-independent* results had been conjectured some time ago [12], and has been recently proven [28] via detailed calculations. Moreover, the PT algorithm gives rise to *exactly* the same answers, regardless of the gauge fixing procedure chosen. This has been shown by explicit calculations for a wide variety of gauge fixing choices, such as the $R_\xi$ gauges, the light-cone gauge [2], the unitary gauge [29], and the background field gauges [25].

The paper is organized as follows: In section 1 we briefly review some of the features of the PT, which are relevant to our purposes. In particular, we present a detailed analysis of the modifications necessary for the application of the PT in the context of the SM with non-conserved external currents. In section 2 the method for constructing the g.i. vertices is described in detail. In section 3 we apply the formalism developed in the previous section to a concrete example, and we perform an explicit one-loop calculation. In section 4 we outline the general method for obtaining WI within the PT framework, and we derive a set of WI for the newly constructed TBVs. In section 5 we explicitly prove the first of the Ward identities derived in the previous section, to one-loop order, in the context of three different gauges. Finally, in section 6 we present our conclusions.

## 2 The Pinch Technique for non conserved currents

The simplest example that demonstrates how the PT works is the gluon two point function (propagator). Consider the $S$-matrix element $T$ for the 2→2 process of the elastic scattering of two fermions of masses $m_1$ and $m_2$ :

$$q_1(p_1) + q_2(p_2) \to q_1(\hat{p}_1) + q_2(\hat{p}_2) \ . \tag{2.1}$$

To any order in perturbation theory $T$ is independent of the gauge fixing parameter one has to use to define the free gluon propagator. For example in the covariant $R_\xi$ gauges the gluon propagator is given by

$$\Delta_{\mu\nu}(q) = \frac{1}{q^2}[g_{\mu\nu} - (1-\xi)\frac{q_\mu q_\nu}{q^2}]. \tag{2.2}$$

On the other hand, as an explicit calculation shows, the conventionally defined proper self-energy depends on the gauge fixing parameter, in this case $\xi$. At the one loop level the gauge dependence of the self-energy graphs is cancelled by contributions from other graphs, vertex or box, which, at first glance, do not seem to be propagator-like. That this cancellation must occur and can be employed to define a g.i. self-energy, is evident from the decomposition:

$$T(s,t,m_1,m_2) = T_1(t,\xi) + T_2(t,m_1,m_2,\xi) + T_3(s,t,m_1,m_2,\xi) \qquad (2.3)$$

where the function $T_1(t)$ depends only on the Mandelstam variable $t = -(\hat{p}_1 - p_1)^2 = -q^2$, and not on $s = (p_1 + p_2)^2$ or on the external masses. Typically, self-energy, vertex, and box diagrams contribute to $T_1$, $T_2$, and $T_3$, respectively. Moreover, such contributions are $\xi$ dependent. However, as the sum $T(s,t,m_1,m_2)$ is g.i., it is easy to show that Eq.(2.3) can be recast in the form

$$T(s,t,m_1,m_2) = \hat{T}_1(t) + \hat{T}_2(t,m_1,m_2) + \hat{T}_3(s,t,m_1,m_2), \qquad (2.4)$$

where the $\hat{T}_i$ ($i = 1,2,3$) are *separately* $\xi$-independent. The propagator-like parts of the vertex and box diagrams which enforce the gauge independence of $T_1(t)$, are called "pinch parts". The pinch parts emerge every time a gluon propagator or an elementary three-gluon vertex contribute a longitudinal $k_\mu$ to the original graph's numerator. The action of such a term is to trigger an elementary Ward identity of the form

$$\begin{aligned} k^\mu \gamma_\mu \equiv \ \slashed{k} &= (\slashed{k} + \slashed{p} - m_i) - (\slashed{p} - m_i) \\ &= S_i^{-1}(p+k) - S_i^{-1}(p) \end{aligned} \qquad (2.5)$$

once it gets contracted with a $\gamma$ matrix. The first term on the right-hand side of 2.5 will remove the internal fermion propagator - that is a "pinch" - whereas $S^{-1}(p)$ vanish on shell. Returning to the decomposition of Eq.(2.4), the function $\hat{T}_1$ is g.i. and may be identified with the contribution of the new propagator. We can construct the new propagator, or equivalently $\hat{T}_1$, directly from the Feynman rules. In doing so it is evident that any value for the gauge parameter $\xi$ may be chosen, since $\hat{T}_1$, $\hat{T}_2$, and $\hat{T}_3$ are all independent of $\xi$. The simplest of all covariant gauges is certainly the Feynman gauge ($\xi = 1$), which removes the longitudinal part of the gluon propagator. Therefore, the only possibility for pinching in four-fermion amplitudes arises from the four-momentum of the three-gluon vertices, and the only propagator-like contributions come from vertex graphs and not from boxes.

The generalization of the PT from vector-like theories (such as QCD) to the SM is technically and conceptually straightforward, as long as one assumes that the external fermionic currents are conserved. For example, applying the PT to a SM amplitude, such as $e^- \nu_e \to e^- \nu_e$, with $m_e = m_\nu = 0$, a $\xi$-independent self-energy for the $W$-boson may be constructed [13].

The situation becomes more involved if one decides to consider non-conserved external fermionic currents, e.g. fermions with non-vanishing masses. The main reasons are the following:

(a) The charged $W$ couples to fermions with different, non-vanishing masses $m_i, m_j \neq 0$, and consequently the elementary Ward identity of Eq.(2.5) gets modified to :

$$k_\mu \gamma^\mu P_L \equiv \slashed{k} P_L = S_i^{-1}(p+k)P_L - P_R S_j^{-1}(p) + m_i P_L - m_j P_R \qquad (2.6)$$

where

$$P_{R,L} = \frac{1 \pm \gamma_5}{2} \qquad (2.7)$$

are the chirality projection operators. The first two terms of Eq.(2.6) will pinch and vanish on shell, respectively, as they did before. But in addition, a term proportional to $m_i P_L - m_j P_R$ is left over. In a general $R_\xi$ gauge such terms give rise to extra propagator and vertex-like contributions, not present in the massless case. For the neutral $Z$ that couples to fermions of the same mass we have to set $m_i = m_j = m$ in Eq.(2.6).

(b) Additional graphs involving the "unphysical" Goldstone bosons $\chi$ and $\phi$, and physical Higgs $H$, which do not couple to massless fermions, must now be included. Such graphs give rise to new pinch contributions, even in the Feynman gauge, due to the momenta carried by interaction vertices such as $\gamma \phi^+ \phi^-$, $Z \phi^+ \phi^-$, $W^+ \phi^- \chi$, $H W^+ \phi^-$ etc, e.g. vertices with one vector gauge boson and two scalar bosons. So, for example, all the graphs of Fig.4 give rise to new vertex-like pinch contributions to the $\gamma W W$ and $Z W W$ vertices, while in the massless case considered in [23], only graphs (1) and $R^{(1,2)}$ were present.

(c) After the pinch contributions have been identified, particular care is needed in deciding how to allot them among the (eventually $\xi$ independent) quantities one is attempting to construct. When constructing g.i. TBVs, for example, in the massless case ($m_i = m_j = 0$), all vertex-like pinch contributions are allotted among the $\gamma W W$ and $Z W W$, the only two vertices which contribute to the amplitude. In the massive case we propose to study, vertices such as $\chi W^- W^+$, $H W^- W^+$, $Z \phi^- W^+$, etc, contribute non-vanishingly to the amplitude, and they must also be rendered g.i. through proper allocation of the available vertex-like pinch parts. The details of how this is accomplished will be presented in the next section.

Before we proceed with the construction of the vertices and the subtleties involved we record some useful formulas. In what follows we use the Feynman rules and the conventions of [30]. The tree-level vector-boson propagator $\Delta^i_{\mu\nu}(q)$ in the $R_\xi$ gauges is given by

$$\Delta^{\mu\nu}_i(q, \xi_i) = \frac{1}{q^2 - M_i^2}[g^{\mu\nu} - (1-\xi_i)\frac{q^\mu q^\nu}{q^2 - \xi_i M_i^2}] \ , \qquad (2.8)$$

with $i = W, Z, \gamma$, and $M_\gamma = 0$. Its inverse $\Delta_i^{-1}(q, \xi_i)^{\mu\nu}$ is given by

$$\Delta_i^{-1}(q, \xi)^{\mu\nu} = (q^2 - M_i^2)g^{\mu\nu} - q^\mu q^\nu + \frac{1}{\xi_i}q^\mu q^\nu \ . \tag{2.9}$$

The propagators $\Delta_s(q, \xi_i)$ of the unphysical (would–be) Goldstone bosons are given by

$$\Delta_s(q, \xi_i) = \frac{-1}{q^2 - \xi_i M_i^2} \ , \tag{2.10}$$

with $(s, i) = (\phi, W)$ or $(\chi, Z)$ and explicitly depend on $\xi_i$. On the other hand, the propagators of the fermions (quarks and leptons), as well as the propagator of the physical Higgs particle are $\xi_i$-independent at tree-level.

The following identities, which hold for *every* value of the gauge fixing parameters $\xi_i$, will be used extensively [21]:

$$\Delta_i^{\mu\nu}(q, \xi_i) = U_i^{\mu\nu}(q) - \frac{q^\mu q^\nu}{M_i^2}\Delta_s(q, \xi_i) \ , \tag{2.11}$$

where

$$U_i^{\mu\nu}(q) = [g^{\mu\nu} - \frac{q^\mu q^\nu}{M_i^2}]\frac{1}{q^2 - M_i^2} \tag{2.12}$$

is the $W$ and $Z$ propagator in the unitary gauge ($\xi_W, \xi_Z \to \infty$) and

$$U_i^{-1}(q)^{\mu\nu} = g^{\mu\nu}(q^2 - M_i^2) - q^\mu q^\nu \tag{2.13}$$

its inverse . Furthermore,

$$g_\nu^\alpha = \Delta_{\nu\mu}^i(q, \xi_i)\Delta_i^{-1}(q, \xi_i)^{\mu\alpha} = \Delta_{\nu\mu}^i(q, \xi_i)U_i^{-1}(q)^{\mu\alpha} - q_\nu q^\alpha \Delta_s(q, \xi_i) \ , \tag{2.14}$$

and

$$q^\mu = -M_i^2 q_\nu \Delta_i^{\nu\mu}(q, \xi_i) - q^2 q^\mu \Delta_s(q, \xi_i). \tag{2.15}$$

Finally, the divergences of the currents $J_Z^\mu$, $J_W^\mu$, $J_W^{+\mu}$ of [13] are related at tree-level to the currents of the would–be Goldstone bosons $J_\chi$, $J_\phi$, $J_\phi^+$ by the following elementary identities:

$$\begin{aligned}
(\bar{e} \chi e) &= \frac{-i q^\mu}{M_Z} (\bar{e} Z e)_\nu \\
(\bar{\nu} \phi^- e) &= \frac{-i p_2^\rho}{M_W} (\bar{\nu} W^- e)_\rho \\
(\bar{e} \phi^+ \nu) &= \frac{-i p_1^\sigma}{M_W} (\bar{\nu} W^+ e)_\sigma
\end{aligned} \tag{2.16}$$

where $q$, $p_1$, and $p_2$ are the momenta carried by the bosons as shown in Fig.2.

## 3 The gauge independent three boson vertices

In this section we show how to use the PT in order to construct gauge invariant three-boson-vertices (TBV), with all three of their incoming momenta *off shell*.

We consider the S-matrix element for the process

$$e^{-}(n) + \nu(\ell) + e^{-}(r) \rightarrow e^{-}(\hat{n}) + e^{-}(\hat{\ell}) + \nu(\hat{r}) \tag{3.1}$$

where

$$q = n - \hat{n} \ , \ p_1 = \ell - \hat{\ell} \ , \ p_2 = r - \hat{r} \ , \tag{3.2}$$

are the momentum transfers at the corresponding fermion lines; they represent the incoming momenta of each of the bosons, merging in the TBV. The TBV's which can be extracted from the S-matrix element of the above process will be in general denoted as $\widehat{\Gamma}^{NLR}$, with $N = \gamma, Z, \chi, H$ ; $L = W^{-}, \phi^{-}$ and $R = W^{+}, \phi^{+}$, where $N$, $L$, and $R$ stand for the neutral, left (positive charge created), and right (positive charge destroyed) legs of the vertex.

We can extract g.i. *improper* vertices by identifying the part $\widehat{T}(q, p_1, p_2)$ of the S-matrix which is independent of the external momenta $n, r, \ell, \hat{n}, \hat{r}, \hat{\ell}$, and only depends on the momentum transfers $q, p_1, p_2$. The general form of $\widehat{T}(q, p_1, p_2)$ is shown in Fig.3. $\widehat{T}(q, p_1, p_2)$ is g.i. as long as we append to the regular vertex graphs all parts of the rest of the graphs, which only depend on the momentum transfers $q, p_1, p_2$. Examples of graphs containing such vertex-like pinch parts are shown in Fig.4.

The inclusion of these extra pieces cancels all $\xi_i$-dependent parts of the regular vertex diagrams; the only gauge dependence remaining stems from the tree-level expressions of the propagators of the boson legs. As we will see in section 5 the cancellation of this residual $\xi_i$ dependence is enforced by a set of WI satisfied by the g.i $\widehat{\Gamma}^{TLR}$'s. The final form of the g.i. $\widehat{T}(q, p_1, p_2)$ is a sum of individually g.i. sub-amplitudes $\widehat{T}^{NLR}(q, p_1, p_2)$ and is given by

$$\widehat{T}(q, p_1, p_2) = \sum_{\{TLR\}} (\bar{e}Te)(\bar{e}L\nu_e)(\bar{\nu}_e Re) \ \widehat{T}^{TLR}(q, p_1, p_2) \tag{3.3}$$

$$= \sum_{\{TLR\}} (\bar{e}Te)(\bar{e}L\nu_e)(\bar{\nu}_e Re) \ \widehat{\Delta}_N(q)\widehat{\Delta}_L(p_1)\widehat{\Delta}_R(p_2) \ \widehat{\Gamma}^{TLR}(q, p_1, p_2)$$

where all internal Lorentz indices have been suppressed. In order to extract the proper $\widehat{\Gamma}^{NLR}(q, p_1, p_2)$ from the respective $\widehat{T}^{NLR}(q, p_1, p_2)$ one must strip off the three g.i. $\widehat{\Delta}$ s, by multiplying $\widehat{T}^{NLR}(q, p_1, p_2)$ with the respective inverse propagators $\widehat{\Delta}^{-1}$. We remind the reader that the $\widehat{\Delta}$ may be individually constructed through the application of the PT to appropriate four-fermion amplitudes (see for example [13] and [16]).

Another equivalent and more economical way to isolate the proper vertex, described in detail in [5] and [23], is to notice that the conventional self-energies of the external boson legs can be converted to the respective g.i. PT self-energies, except for certain missing pinch pieces. These missing pieces may be supplemented to the self-energy by hand, and correspondingly subtracted from the TBV. All such terms are multiplied by an inverse tree-level propagator (which is the characteristic structure of all pinch terms), and they remove the tree-level boson propagator connecting them to the rest of the graph. Therefore, they are effectively one-particle *irreducible*, and they may be freely added to the rest of the one-particle irreducible terms contributing to the TBV [6].

Schematically the g.i. $TBV$ $\widehat{\Gamma}_{NLR}$ will consist of the following pieces :

$$\widehat{\Gamma}_{NLR} = \Gamma_{NLR}^{(\xi_i=1)} + \Gamma_{NLR}^{P} - \frac{1}{2}\Pi_{NN'}^{P}\Gamma_{N'LR}^{(0)} - \frac{1}{2}\Pi_{LL'}^{P}\Gamma_{NL'R}^{(0)} - \frac{1}{2}\Pi_{RR'}^{P}\Gamma_{NLR'}^{(0)} \qquad (3.4)$$

where $\Gamma_{NLR}^{(\xi_i=1)}$ are the conventional graphs contributing to the $TBV$ in the Feynman gauge, $\Gamma_{NLR}^{P}$ are all vertex-like pinch parts of box diagrams (also computed with $\xi_i = 1$ [31] ), which are kinematically equivalent to the TBV in question (this point will be further clarified later in this section), $\Gamma_{NLR}^{(0)}$ are tree-level expressions of respective TBVs, and $\Pi_{ij}^{P}$ ($i,j = N,L,R$) is the pinch contribution to the $ij$-boson self-energy (again at $\xi_i = 1$). Since the derivation of the pinch parts of propagators has been extensively discussed in the literature, we will first focus on the technical details pertaining to the construction of the term $\Gamma_{NLR}^{P}$ in Eq.(3.4)

The pinch parts of graphs are extracted using Eq.(2.6), whenever possible. The box diagrams of Fig.4 for example, represent the complete set of diagrams that can contribute vertex-like parts to the $\gamma W^- W^+$, $ZW^- W^+$, and $\chi W^- W^+$ vertices [32]. Depending on which of the internal fermion propagator has been removed, the vertex-like pinch amplitudes assume one of the following forms:

$$(\bar{e}L\nu)...(\bar{\nu}Re)...\Delta^L_{...}(p_1)\Delta^R_{...}(p_2) \left[ \frac{g^2}{2}(\bar{e}\gamma^\mu P_L e)\, B^N_{...\mu} \;+\; \frac{g^2}{2}m_e(\bar{e}P_L e)\, \mathcal{M}^N_{...} \right]$$

$$(\bar{\nu}Re)...(\bar{e}Ne)...\Delta^R_{...}(p_2)\Delta^N_{...}(q) \left[ (\bar{e}W^+\nu)^\mu\, B^L_{...\mu} \;+\; iM_W(\bar{e}\phi^+\nu)\, \mathcal{M}^L_{...} \right]$$

$$(\bar{e}Ne)...(\bar{e}L\nu)...\Delta^N_{...}(q)\Delta^L_{...}(p_1) \left[ (\bar{\nu}W^- e)^\mu\, B^R_{...\mu} \;+\; iM_W(\bar{\nu}\phi^- e)\, \mathcal{M}^R_{...} \right] \qquad (3.5)$$

The ellipses in Eq.(3.5) represent appropriately contracted Lorentz indices, which we suppress. We note that the $\mathcal{M}$ terms originate from the mass left-overs of Eq.(2.6). The factors $B$ and $\mathcal{M}$ in the expressions above are in general complicated functions of $q$, $p_1$, and $p_2$, and boson masses; however they do not depend on the individual momenta and masses of the external "test" fermions. Notice also that they are ultra-violet finite since they originated from box diagrams. Clearly, the $B$'s or the $\mathcal{M}$'s may be zero for some graphs. Once all relevant pinch contributions have been extracted, they must be judiciously allotted to the appropriate TBV's. To that end, one has to perform the following three steps:

(i) The couplings multiplying the $B$ and $\mathcal{M}$ in the r.h.s. of the first relation in Eq.(3.5) must be rewritten as a linear combination of the couplings of the bosons which can be attached to the corresponding fermion current. So, for the couplings of the neutral bosons on the top fermion line we write :

$$\frac{g^2}{2}\gamma^\rho P_L = -gc(\bar{e}Ze)^\rho - gs(\bar{e}\gamma e)^\rho$$
$$\frac{g^2}{2}m_e P_L = -gM_W(\bar{e}He) + gc\frac{iM_Z}{2}(\bar{e}\chi e) \qquad (3.6)$$

On the other hand, the appropriate couplings for the charged bosons have already appeared in the r.h.s. of the second and third relation of Eq.(3.5).

(ii) We use the identities given in Eq.(2.17) to rewrite the couplings of the Goldstone bosons to the fermions as divergences of the corresponding currents of the gauge bosons. At the end of these two steps the pinch parts in the square brackets of Eq.(3.5) assume the following form:

$$\left[-gc(\bar{e}Ze)^\rho \left(B^N_{\cdots\rho} - \frac{q_\rho}{2}\mathcal{M}^N_{\cdots}\right) - gs(\bar{e}\gamma e)^\rho B^N_{\cdots\mu} - gM_W(\bar{e}He)\mathcal{M}^N_{\cdots}\right] \qquad (3.7)$$

$$\left[B^L_{\cdots\rho} + p_{1\rho}\mathcal{M}^L_{\cdots}\right](\bar{e}W^+\nu)^\rho \qquad (3.8)$$

$$\left[B^R_{\cdots\rho} + p_{2\rho}\mathcal{M}^R_{\cdots}\right](\bar{\nu}W^-e)^\rho \qquad (3.9)$$

(iii) The final step in transforming these expressions into the desired form of TBV is to recognize that a tree-level boson propagator must be attached at the point where the pinching took place. It is straightforward to make the missing photon and Higgs propagator appear. We only need to insert unity written as a product of a propagator and its inverse. The inverse propagator will be incorporated to the rest of the pinch expression. We emphasize that no additional $\xi$ dependencies are introduced into the pinch expressions through this process, since the part of the inverse photon propagator proportional to $\xi_\gamma$ vanishes from the amplitude due to conservation of the electromagnetic current $J^\mu_\gamma$, whereas the Higgs propagator and its inverse are g.i at tree-level. In order to accomplish this last step for the massive gauge bosons, we have to use the identities of Eq.(2.14), since now the relevant currents $J^\mu_Z$ and $J^\mu_W$ are not conserved. Finally we obtain :

$$(\bar{e}\gamma e)_\rho \frac{g^{\rho\nu}}{q^2}\left[-gs\ T^{\nu\mu}(q)\ B^N_{\cdots\mu}\right] + (\bar{e}He)\ \Delta_H(q)\left[-gM_W\Delta_H^{-1}(q)\mathcal{M}^N_{\cdots}\right]$$

$$+ (\bar{e}Ze)_\rho \, \Delta_Z^{\rho\nu}(q) \left[ -gcU_{\nu\mu}^{-1}(q)_Z \left( B_{\ldots\mu}^N + \frac{q^\mu}{2}\mathcal{M}_{\ldots}^N \right) \right]$$

$$+ (\bar{e}\chi e) \, \Delta_\chi(q) \left[ \, iM_Z gcq^\mu \left( B_{\ldots\mu}^N + \frac{q^\mu}{2}\mathcal{M}_{\ldots}^N \right) \, \right] \qquad (3.10)$$

$$\left[ B_{\ldots\mu}^L + p_{1\mu}\mathcal{M}_{\ldots}^L \right] \left[ U_W^{-1}(p_1)_{\nu\mu} \Delta_W^{\rho\nu}(p_1,\xi_W)(\bar{e}W^+\nu)_\rho - iM_W p_1^\mu \Delta_\phi(p_1,\xi_W)(\bar{e}\phi^+\nu) \right] \quad (3.11)$$

$$\left[ B_{\ldots\mu}^R + p_{2\mu}\mathcal{M}_{\ldots}^R \right] \left[ U_W^{-1}(p_2)_{\nu\mu} \Delta_W^{\rho\nu}(p_2,\xi_W)(\bar{e}W^-\nu)_\rho - iM_W p_2^\mu \Delta_\phi(p_2,\xi_W)(\bar{e}\phi^-\nu) \right] \quad (3.12)$$

It is now evident how the pinch parts must be allotted among the various (eventually g.i.) TVB's. We demonstrate it schematically below.

From the graphs that pinch at the top (neutral) fermion line (Eq.3.10) the pinch parts are distributed as follows ;

$$\begin{aligned}
-gs\, T^{\nu\mu}(q)\, B_{\ldots\mu}^N &\longrightarrow \widehat{\Gamma}_{\nu\ldots}^{\gamma LR} \\
-gcU_Z^{-1}(q)_{\nu\mu}\left( B_{\ldots\mu}^N + \frac{q^\mu}{2}\mathcal{M}_{\ldots}^N \right) &\longrightarrow \widehat{\Gamma}_{\nu\ldots}^{ZLR} \\
iM_Z gcq^\mu \left( B_{\ldots\mu}^N + \frac{q^\mu}{2}\mathcal{M}_{\ldots}^N \right) &\longrightarrow \widehat{\Gamma}^{\chi LR} \\
-gM_W \Delta_H^{-1}(q)\mathcal{M}_{\ldots}^N &\longrightarrow \widehat{\Gamma}^{HLR}
\end{aligned} \qquad (3.13)$$

From the graphs that pinch on the left (Eq.3.11) we have :

$$\begin{aligned}
U_W^{-1}(p_1)_{\nu\mu}\left( B_{\ldots\mu}^L - p_{1\mu}\mathcal{M}_{\ldots}^L \right) &\longrightarrow \widehat{\Gamma}_{\cdot\nu\cdot}^{NW^-R} \\
iM_W p_1^\mu \left( B_{\ldots\mu}^L - p_{1\mu}\mathcal{M}_{\ldots}^L \right) &\longrightarrow \widehat{\Gamma}^{N\phi^-R}
\end{aligned} \qquad (3.14)$$

and from the graphs that pinch on the right (Eq.3.12) :

$$\begin{aligned}
U_W^{-1}(p_2)_{\nu\mu}\left( B_{\ldots\mu}^R - p_{2\mu}\mathcal{M}_{\ldots}^R \right) &\longrightarrow \widehat{\Gamma}_{\cdot\cdot\nu}^{NLW^+} \\
iM_W p_2^\mu \left( B_{\ldots\mu}^R - p_{2\mu}\mathcal{M}_{\ldots}^R \right) &\longrightarrow \widehat{\Gamma}^{NL\phi^+}
\end{aligned} \qquad (3.15)$$

The final step in the construction of the g.i. TBVs is the inclusion of all pinch terms that have been left over from converting gauge dependent boson self energies into their gauge independent PT counterparts, at other parts of the amplitude considered; they constitute the third term in the r.h.s. of Eq.(3.4). To begin with, it is important to recognize that in addition to the boson legs attached to the TBVs, the boson legs of the "monodromic" graphs (collectively depicted in Fig.8 (e),(f),(g)) must be rendered g.i. We call them "monodromic" (one-way), because their graph structure of vertices and edges (propagators) is that of an "Eulerian-path" or self avoiding curve. That is , all the vertices can be visited by a line that does not run through an edge twice. Notice that they contain an off-shell fermion propagator. In the rest of this section we outline how such pieces are included in the vertices through a specific example.

Let us concentrate on the $Z$ self–energy legs. In the Feynman gauge ($\xi_i = 1$), the only propagator-like pinch parts for the g.i. $Z$ self-energy originate from the graph shown in Fig.3.a and its mirror graph Fig.3.b and their contribution is equal. For the $Z$ self–energy leg attached to the $ZWW$ vertex, one of the above graphs (Fig.3.c), is already present in the amplitude we consider and supplies half of the necessary pinch contribution. The other half, where the pinch would occur at the side where we now have the TBV vertex, is missing. Therefore its pinch contribution must be supplemented by hand to the $Z$ self-energy graphs and subsequently subtracted from the $ZWW$ vertex graphs. We observe that this contribution to the vertex will be of the form $-2g^2c^2 I_{WW}(q) \left[U_Z^{\mu\nu}(q)\right]^{-1} \Delta_Z^{\nu\rho}(q,\xi_Z) \Gamma_{\rho\alpha\beta}^{ZWW}$. This last expression is explicitly gauge dependent. The effect of the monodromic graphs is to precisely cancel this residual gauge dependence. To understand how this cancellation mechanism works we now concentrate on the vertex–like pieces originating from the monodromic graphs. As a first step, their bosonic legs must be rendered g.i; in doing so we notice that, unlike the previous case, all the necessary propagator–like pinch parts are now available (an example of a graph that contributes such a pinch term is shown in Fig.3(d)). One then proceeds as usually and first pinches the fermion propagator inside the loop and then uses Eqs.(2.14 , 2.15) to attach boson propagators at the point where the pinching took place. At this point one observes that the momenta accompanying the part with the scalar propagator $\Delta_s$ in Eqs.(2.14 , 2.15) can trigger additional pinching and remove the remaining fermion propagator that was outside of the loop. Thus a vertex-like piece finally emerges from this part and must be included with the rest of the vertex graphs. Clearly, all these pieces are also explicitly gauge dependent since they carry a $\Delta_\chi(q,\xi_Z)$, and by using Eq.(2.11) exactly cancel against the relevant $\Delta_\chi(q,\xi_Z)$ part coming from the leg attached to the TBV. In the remaining expression the tree level propagators in the unitary gauge also cancel and the part that needs to be appended to the $ZWW$ vertex is $-2g^2c^2 I_{WW}(q) \Gamma_{\mu\alpha\beta}^{ZWW}$. A similar procedure must be followed case by case for all the TBVs and will conclude the construction of a g.i. three boson vertex.

## 4 The vertices $\gamma W^-W^+$, $ZW^-W^+$, $\chi W^-W^+$

In the previous section we presented the general procedure for constructing g.i. TBVs via the PT. In this section we focus on three particular TBVs, namely $\widehat{\Gamma}^{\gamma W^-W^+}$, $\widehat{\Gamma}^{ZW^-W^+}$ and $\widehat{\Gamma}^{\chi W^-W^+}$, and we describe in detail their derivation. This section is rather technical; we present several intermediate results, which will also be used in subsequent sections.

The final expressions for $\widehat{\Gamma}_{\mu\alpha\beta}^{\gamma W^-W^+}$, $\widehat{\Gamma}_{\mu\alpha\beta}^{ZW^-W^+}$ and $\widehat{\Gamma}_{\alpha\beta}^{\chi W^-W^+}$, are summarised in Eq.(4.33)–Eq.(4.35).

We adopt the following convention: The scalar parts of boson propagators of mass $M_A$ and momentum $q$ will be denoted by

$$A(q) \equiv \frac{1}{q^2 - M_A^2} \, . \tag{4.1}$$

For example, with this notation the tree-level propagator for the $W$ in Eq.(2.8) assumes the form :

$$\Delta_W^{\mu\nu}(q) = [g^{\mu\nu} - (1-\xi_i)\frac{q^\mu q^\nu}{q^2 - \xi_i M_i^2}] \, W(q) \, , \tag{4.2}$$

We introduce the following short-hand notation:

$$\int (ABC) \, \{...\} \equiv \int (dk) A(k+p_1) B(k-p_2) C(k) \, \{...\} \, , \tag{4.3}$$

where the momentum integration measure is $(dk) = \frac{d^4 k}{i(2\pi)^4}$ for convergent integrals and $(dk) = \mu^{4-n} \frac{d^n k}{i(2\pi)^n}$ for dimensionally regularized integrals. Furthermore, we define the scalar integrals

$$J_{ABC} \equiv J_{ABC}(q, p_1, p_2) = \int (ABC) \, , \tag{4.4}$$

$$I_{AB}(q) = \int (dk) A(k) B(k+q) \, . \tag{4.5}$$

The box diagrams containing vertex-like contributions, in the Feynman gauge are shown in Fig.4. From the first two diagrams of Fig.4, which we treat as one, we obtain

$$N_{\mu\alpha\beta}^1 = gc \, U_Z^{-1}(q)_\mu^\rho \, g^2 B_{\rho\alpha\beta} \tag{4.6}$$

and

$$\mathcal{N}_{\alpha\beta}^1 = -iM_Z \, gc \, q^\rho g^2 B_{\rho\alpha\beta} \, , \tag{4.7}$$

where $g^2 B_{\rho\alpha\beta}$ is the same expression as in the case for conserved currents (see [23] Eq.3.5 and Eq.3.6), namely

$$g^2 B_{\rho\alpha\beta} = \sum_{V=\gamma,Z} g_V^2 \int (WWV) \left[ g_{\alpha\beta} \left(k - \frac{3}{2}(p_1 - p_2)\right)_\mu - g_{\alpha\mu}(3k+2q)_\beta - g_{\beta\mu}(3k-2q)_\alpha \right] \, . \tag{4.8}$$

with $g_\gamma = gs$ and $g_Z = gc$. $N_{\mu\alpha\beta}^1$ is allotted to the vertex $\widehat{\Gamma}_{\mu\alpha\beta}^{ZW^-W^+}$ whereas $\mathcal{N}_{\alpha\beta}^1$ to $\widehat{\Gamma}_{\alpha\beta}^{\chi W^-W^+}$.

Similarly, the graphs containing a Higgs boson (2 and 3 in Fig.4) yield:

$$N_{\mu\alpha\beta}^2 + N_{\mu\alpha\beta}^3 = -M_Z^2 \, g^3 c \, q_\mu g_{\alpha\beta} \, \mathcal{M} \, , \tag{4.9}$$

and
$$\mathcal{N}_{\alpha\beta}^2 + \mathcal{N}_{\alpha\beta}^3 = -iM_Z\, q^2\, g^3 c\, g_{\alpha\beta}\, \mathcal{M}\,,\qquad(4.10)$$

with
$$\mathcal{M} = \frac{1}{2}\left(J_{HZW} + J_{ZHW}\right).\qquad(4.11)$$

We note that box diagrams which contain any two internal neutral bosons, except the Higgs boson, give zero total pinch contribution. This is so because the pinch parts of the direct diagrams cancel against the corresponding pinch parts of the crossed diagram. Similarly, the pinch contributions of diagrams with one $\phi$ and one $W$ in the loop cancel against the corresponding contribution from the mirror graphs, e.g. $W \leftrightarrow \phi$.

The pinch contributions of the diagrams 1,...,6 of the second row of Fig.4, where the pinching occurs at the leg of the $W^+$, are extracted following exactly similar steps. We denote these pinch contributions by $R^i_{\mu\alpha\beta}$ where $i = 1,..,6$. In what follows the suffix $cr$ is used to denote the inclusion of the crossed graphs which are not shown in Fig.4. The relevant expressions of the pinch contributions of these graphs are :

$$R_{\mu\alpha\beta}^{(1,2,2cr)} = gc\, U_W^{-1}(p_2)_\beta^\rho g^2 B_{\mu\alpha\rho}^- + R_{\mu\alpha\beta}^1 + R_{\mu\alpha\beta}^2\,,\qquad(4.12)$$

where
$$R_{\mu\alpha\beta}^1 = g^3 cs^2 M_W^2\, p_{2\beta} g_{\mu\alpha} J_{WW\gamma}\,,\quad R_{\mu\alpha\beta}^2 = g^3 c\left(\frac{1-2s^2}{2}\right) M_W^2\, p_{2\beta} g_{\mu\alpha} J_{WWZ}\qquad(4.13)$$

and
$$g^2 B_{\mu\alpha\beta}^-(q,p_1,p_2) = \sum_{V=\gamma,Z} g_V^2 \int (WWV)\, G_{\mu\alpha\beta}(q,p_1,p_2)\,,\qquad(4.14)$$

with
$$G_{\mu\alpha\beta}(q,p_1,p_2) = g_{\alpha\beta}\left(3k + 3p_1 - 2p_2\right)_\mu + g_{\mu\beta}\left(3k + p_1 - 2q\right)_\alpha - g_{\alpha\mu}\left(k + 2p_1 - 2q\right)_\beta\,.\qquad(4.15)$$

Notice that the result of the conserved current case can be recovered from Eq.(4.14) if we neglect the terms proportional to $p_{1\alpha}$ and $p_{2\beta}$ (Eq.3.11 of [23]).

The rest of the diagrams give :

$$R_{\mu\alpha\beta}^3 = g^3 \frac{s^4}{c} M_W^2\, p_{2\beta} g_{\mu\alpha} J_{WW\gamma}\,,\qquad(4.16)$$

$$R_{\mu\alpha\beta}^4 + R_{\mu\alpha\beta}^{4cr} = g^3 \frac{s^2(1-2s^2)}{2c} M_W^2\, p_{2\beta} g_{\mu\alpha}\,, J_{WWZ}\qquad(4.17)$$

$$R_{\mu\alpha\beta}^5 = \frac{g^3 c M_W^2}{2}\, p_{2\beta} g_{\mu\alpha} J_{WWH}\,,\qquad(4.18)$$

$$R^6_{\mu\alpha\beta} = \frac{g^3 M_W^2}{2c} p_{2\beta} g_{\mu\alpha} J_{ZHW} , \qquad (4.19)$$

We next turn to the rest of the graphs of Fig.4, third and fourth row, and isolate the pinch contributions, which will be appended to the vertex $\widehat{\Gamma}^{\chi W^- W^+}_{\alpha\beta}$. We denote them by $\mathcal{R}^i_{\alpha\beta}$ where $i = 1,..9$. Their explicit expressions are:

$$\mathcal{R}^1_{\alpha\beta} + \mathcal{R}^2_{\alpha\beta} + \mathcal{R}^{2cr}_{\alpha\beta} = \frac{g^3}{iM_Z}\frac{s^2}{2c} M_W^2 U_W^{-1}(p_2)_{\alpha\beta} \sum_{V=\gamma,Z} b_V J_{WWV} , \qquad (4.20)$$

$$\mathcal{R}^3_{\alpha\beta} = \frac{g^3}{iM_Z}\frac{s^2}{2c} M_W^2 p_{2\beta} \int (WW\gamma)(k-2q)_\alpha , \qquad (4.21)$$

$$\mathcal{R}^4_{\alpha\beta} + \mathcal{R}^{4cr}_{\alpha\beta} + \mathcal{R}^5_{\alpha\beta} = \frac{g^3}{iM_Z}\frac{cM_W^2}{2} p_{2\beta} \int (WWZ) k_\alpha \qquad (4.22)$$
$$- \frac{g^3}{iM_Z}\frac{1-s^2}{2c} M_W^2 p_{2\beta} q_\alpha J_{WWZ} + \frac{g^3 M_W^2}{i8cM_Z} p_{1\alpha} p_{2\beta} J_{WWZ} .$$

The last term in the r.h.s. of Eq.(4.22) cancels against the corresponding contribution from the left, coming from the graphs $\mathcal{L}^4_{\alpha\beta} + \mathcal{L}^{4cr}_{\alpha\beta} + \mathcal{L}^5_{\alpha\beta}$.

$$\mathcal{R}^6_{\alpha\beta} = \frac{g^3}{iM_Z}\frac{M_W^2}{4c} p_{2\beta} \int (WWH) k_\alpha , \qquad (4.23)$$

$$\mathcal{R}^7_{\alpha\beta} = -\frac{g^3}{iM_Z}\frac{M_W^2}{2c} q_\alpha p_{2\beta} J_{ZHW} , \qquad (4.24)$$

$$\mathcal{R}^8_{\alpha\beta} + \mathcal{R}^{8cr}_{\alpha\beta} = \frac{g^3}{iM_Z}\frac{M_W^2}{2c} U_W^{-1}(p_2)_{\alpha\beta} J_{HZW} , \qquad (4.25)$$

$$\mathcal{R}^9_{\alpha\beta} + \mathcal{R}^{9cr}_{\alpha\beta} = \frac{g^3}{iM_Z}\frac{(1-2s^2)}{4c} M_Z^2 p_{2\beta} \int (HZW)(2k+p_1)_\alpha . \qquad (4.26)$$

The corresponding diagrams where the pinching occurs at the left fermion line we will denote as $L^i_{\mu\rho\beta}$ and $\mathcal{L}^i_{\rho\beta}$ respectively. (these diagrams are not shown). They are given by:

$$L^i_{\mu\alpha\beta}(q,p_1,p_2) = - R^i_{\mu\beta\alpha}(q,p_2,p_1) , \quad \mathcal{L}^i_{\alpha\beta}(q,p_1,p_2) = - \mathcal{R}^i_{\beta\alpha}(q,p_2,p_1) \qquad (4.27)$$

The total pinch contribution is the sum of all relevant terms. We define :

$$\sum_{i=1}^{6} R^i_{\mu\alpha\beta} = g^3 c M_W^2 p_{2\beta} g_{\mu\alpha} \mathcal{M}^- , \qquad \sum_{i=1}^{6} L^i_{\mu\alpha\beta} = g^3 c M_W^2 p_{1\alpha} g_{\mu\beta} \mathcal{M}^+ , \qquad (4.28)$$

$$\sum_{i=1}^{9} \mathcal{R}^i_{\alpha\beta} = \frac{g^3 c}{iM_Z} \mathcal{R}^-_{\alpha\beta} , \qquad \sum_{i=1}^{6} \mathcal{L}^i_{\alpha\beta} = \frac{g^3 c}{iM_Z} \mathcal{L}^+_{\alpha\beta} \qquad (4.29)$$

where crossed graphs are included in the sums and

$$\mathcal{M}^-(q,p_1,p_2) = \frac{s^2}{c^2}J_{WW\gamma} + \frac{1-2s^2}{2c^2}J_{WWZ} + \frac{1}{2}J_{WWH} + \frac{1}{2c^2}J_{ZHW} \; , \tag{4.30}$$

$$\mathcal{M}^+(q,p_1,p_2) = -\mathcal{M}^-(q,p_2,p_1) \; . \tag{4.31}$$

The last step is to add the pinch contributions to the regular vertex graphs. Thus, if we define

$$g^2 g \Gamma^{VW^-W^+}_{\mu\alpha\beta}|_{\xi_i=1} = \sum_{i=1}^{n_V} V^i_{\mu\alpha\beta}|_{\xi_i=1} \; , \quad g^3 c \Gamma^{\chi W^-W^+}_{\alpha\beta}|_{\xi_i=1} = \sum_{i=1}^{21} \mathcal{S}^i_{\alpha\beta}|_{\xi_i=1} \; , \tag{4.32}$$

to be the sum of the usual graphs of the respective vertices in the Feynman gauge (depicted respectively at Fig.5, with V= $\gamma, Z$, $n_\gamma = 28$, $n_Z = 34$ and Fig.6), we arrive at the following expressions for the g.i. TBVs

$$\begin{aligned}
\frac{1}{g^3 s}\widehat{\Gamma}^{\gamma W^-W^+}_{\mu\alpha\beta} &= \Gamma^{\gamma W^-W^+}_{\mu\alpha\beta}|_{\xi_i=1} + q^2 T(q)^\rho_\mu B_{\rho\alpha\beta} + U_W^{-1}(p_1)^\rho_\alpha B^+_{\mu\rho\beta} + U_W^{-1}(p_2)^\rho_\beta B^-_{\mu\alpha\rho} \\
&\quad - 2\Gamma_{\mu\alpha\beta}\left[\, I_{WW}(q) + s^2 I_{W\gamma}(p_1) + c^2 I_{WZ}(p_1) + s^2 I_{W\gamma}(p_2) + c^2 I_{WZ}(p_2) \,\right] \\
&\quad + p_{2\beta} g_{\mu\alpha} \mathcal{M}^- + p_{1\alpha} g_{\mu\beta} \mathcal{M}^+ \; ,
\end{aligned} \tag{4.33}$$

$$\begin{aligned}
\frac{1}{g^3 c}\widehat{\Gamma}^{ZW^-W^+}_{\mu\alpha\beta} &= \Gamma^{ZW^-W^+}_{\mu\alpha\beta}|_{\xi_i=1} + U_Z^{-1}(q)^\rho_\mu B_{\rho\alpha\beta} + U_W^{-1}(p_1)^\rho_\alpha B^+_{\mu\rho\beta} + U_W^{-1}(p_2)^\rho_\beta B^-_{\mu\alpha\rho} \\
&\quad - 2\Gamma_{\mu\alpha\beta}\left[\, I_{WW}(q) + s^2 I_{W\gamma}(p_1) + c^2 I_{WZ}(p_1) + s^2 I_{W\gamma}(p_2) + c^2 I_{WZ}(p_2) \,\right] \\
&\quad + q_\mu g_{\alpha\beta} M_Z^2 \mathcal{M} + p_{2\beta} g_{\mu\alpha} M_W^2 \mathcal{M}^- + p_{1\alpha} g_{\mu\beta} M_W^2 \mathcal{M}^+ \; ,
\end{aligned} \tag{4.34}$$

$$\frac{1}{g^3 c}\widehat{\Gamma}^{\chi W^-W^+}_{\alpha\beta} = \Gamma^{\chi W^-W^+}_{\alpha\beta}|_{\xi_i=1} - iM_Z q^\rho B_{\rho\alpha\beta} - iM_Z q^2 \mathcal{M} - \frac{i}{M_Z}\mathcal{R}^-_{\alpha\beta} - \frac{i}{M_Z}\mathcal{L}^+_{\alpha\beta} \; . \tag{4.35}$$

## 5  The Ward Identities

In the previous two sections we outlined the construction of a generic g.i. TBV, and we computed the exact one-loop closed forms for the g.i. $\widehat{\Gamma}^{\gamma W^-W^+}_{\mu\alpha\beta}$, $\widehat{\Gamma}^{ZW^-W^+}_{\mu\alpha\beta}$, $\widehat{\Gamma}^{\chi W^-W^+}_{\alpha\beta}$. In this section we proceed to derive a set of Ward identities that the g.i. TBV satisfy. These Ward identities are a direct consequence of the gauge independence of the S-matrix, order by order in perturbation theory. It should be emphasized that the derivation of the WI does not require knowledge of the explicit closed form of the TBVs involved.

After the construction of g.i. TBVs has been completed, the amplitude we consider has been reorganized into individually $\xi$-independent structures connected by $\xi$-dependent tree level propagators. In other words, the PT algorithm only cancels all $\xi$-dependencies originating from the tree-level propagators appearing *inside* the loops, but a residual $\xi$-dependence, stemming from boson propagators outside of loops, survives at the end of the pinching process. The cancellation of this last $\xi$-dependence becomes possible due to a set of WI satisfied by the g.i. TBV. One can actually derive these WI *without* any detailed knowledge of the algorithm which gives rise to the g.i. TBV. All one needs to assume is that such an algorithm exists (in our case the PT algorithm), and that all residual $\xi$-dependencies should cancel from the S-matrix. So, once the g.i. TBVs have been constructed, one should examine whether or not they actually satisfy the required WI, as a self-consistency check. In this section we use the above arguments to derive the WIs, and we will explicitly check their validity at one-loop in the next section.

It is instructive to illustrate the derivation of WI for a simpler case, namely the g.i. $W$ propagator. We consider the one-loop S-matrix element of the process

$$e^-(b) + \nu_e(t) \to \nu_e(\hat{b}) + e^-(\hat{t}) \tag{5.1}$$

with $q = t - \hat{t} = \hat{b} - b$, and apply the PT rules. As shown in Fig.7, the part of the S-matrix which only depends on $q^2$ assumes the form:

$$\hat{T}_1 = (\bar{e}W^+\nu)_\rho \, \Delta_W^{\rho\mu} \, \widehat{\Pi}_{\mu\nu}^W \Delta_W^{\nu\sigma} \, (\bar{\nu}W^-e)_\sigma + (\bar{e}W^+\nu)_\rho \, \Delta_W^{\rho\mu} \, \widehat{\Pi}_\mu^+ \, \Delta_\phi \, (\bar{\nu}\phi^-e)$$
$$+ (\bar{e}\phi^+\nu) \, \Delta_\phi \, \widehat{\Pi}_\nu^- \, \Delta_W^{\nu\sigma} \, (\bar{\nu}W^-e)_\sigma + (\bar{e}\phi^+\nu) \, \Delta_\phi \, \widehat{\Pi}^\phi \, \Delta_\phi \, (\bar{\nu}\phi^-e) \tag{5.2}$$

Using Eq.(2.17) in order to pull out the factor $(\bar{e}W^+\nu)_\rho \, (\bar{\nu}W^-e)_\sigma$, as well as Eq.(2.14), we can cast the above expression in the following form :

$$\hat{T}_1 = (\bar{e}W^+\nu)_\rho \left[ \left( U_W^{\rho\mu} - \frac{q^\rho q^\mu}{M_W^2}\Delta_\phi \right) \widehat{\Pi}_{\mu\nu}^W \left( U_W^{\nu\sigma} - \frac{q_\nu q_\sigma}{M_W^2}\Delta_\phi \right) + \frac{(-iq^\rho)}{M_W}\Delta_\phi\widehat{\Pi}^\phi\Delta_\phi\frac{iq_\sigma}{M_W} \right.$$
$$\left. + \left( U_W^{\rho\mu} - \frac{q_\rho q_\mu}{M_W^2}\Delta_\phi \right) \widehat{\Pi}_\mu^+ \Delta_\phi\frac{iq_\sigma}{M_W} + \frac{(-iq^\rho)}{M_W}\Delta_\phi\widehat{\Pi}_{\mu\nu}^W \left( U_W^{\nu\sigma} - \frac{q_\nu q_\sigma}{M_W^2}\Delta_\phi \right) \right] (\bar{\nu}W^-e)_\sigma \tag{5.3}$$

In this last expression the $\xi$-dependence is carried solely by the tree-level Goldstone boson propagators $\Delta_\phi(q,\xi_W)$. The requirement that $\hat{T}_1$ is $\xi$-independent, gives rise to two independent equations; the first enforces the cancellation of the terms with only one $\Delta_\phi$ factor, whereas the second enforces the cancellation of the terms with a $\Delta_\phi \Delta_\phi$ factor. It is then a matter of simple algebra to show that the following WI should hold [33] :

$$q^\mu \widehat{\Pi}_{\mu\nu}^W(q) \mp iM_W \widehat{\Pi}_\nu^\pm(q) = 0 \tag{5.4}$$

$$q^\mu \widehat{\Pi}_\mu^\pm(q) \pm iM_W \widehat{\Pi}^\phi(q) = 0 \tag{5.5}$$

$$q^\mu q^\nu \widehat{\Pi}^W_{\mu\nu}(q) - M_W^2 \widehat{\Pi}^\phi(q) = 0 \tag{5.6}$$

Similarly, the requirement of gauge independence for the S-matrix element of a neutral current process gives rise to the following set of WI, relating the two-point Greens functions of $Z$ and its Goldstone boson $\chi$ :

$$q^\mu \widehat{\Pi}^Z_{\mu\nu}(q) - iM_Z \widehat{\Pi}^{Z\chi}_\nu(q) = 0 \tag{5.7}$$

$$q^\mu \widehat{\Pi}^{Z\chi}_\mu(q) + iM_Z \widehat{\Pi}^\chi(q) = 0 \tag{5.8}$$

$$q^\mu q^\nu \widehat{\Pi}^Z_{\mu\nu}(q) - M_Z^2 \widehat{\Pi}^\chi(q) = 0 \tag{5.9}$$

We now turn to our main objective, namely the derivation of the WI for the g.i TBVs. We consider again the S-matrix element of the process in Eq.(3.1). After the pinching is performed we focus on the diagrams of Fig.8, where now the "blobs" represent g.i. expressions. As before the residual $\xi$-dependence of these graphs enters only through the tree-level bosonic propagators (solid, not-oriented lines) We call these graphs respectively :

(i) three-boson vertex graphs ( Fig.8 (a) )
(ii) self-energy graphs ( Fig.8 (b),(c),(d) )
(iii) monodromic graphs ( Fig.8 (e),(f),(g) ) [34]

At first sight, the monodromic graphs do not appear to be akin to the graphs of type (i) and (ii) (which only depend on the momentum transfers $q, p_1, p_2$), since they seem to explicitly depend on the external fermion momenta $n, l, r$ or $\hat{n}, \hat{l}, \hat{r}$, through the internal off-shell fermion propagators. Equivalently, one might think that the characteristic factor $(\bar{e}Te)(\bar{e}L\nu_e)(\bar{\nu}_eRe)$, containing the external fermionic currents, cannot be pulled out from the monodromic graphs. One should notice however, that, in the monodromic graphs additional pinching can take place, triggered by the longitudinal part of the bare vector boson propagators, thus eliminating the dependence on the internal fermion propagator. These pinch parts are vertex-like, and will therefore combine with the graphs of (i) and (ii), in order to cancel the remaining gauge dependence from the amplitude.

To demon¡strate this final cancellation, we use again Eq.(2.11) in order to isolate the residual gauge dependence of the S-matrix into bare Goldstone boson propagators only. All gauge dependent terms will display a characteristic structure, depending on the number and kind of Goldstone boson propagators they contain, and the momenta they carry. Clearly, all such terms form linearly independent combinations. A term with a gauge dependence of the form $\Delta_\chi(q, \xi_Z)\Delta_\phi(p_1, \xi_W)$, for example, cannot cancel against a term of the form $\Delta_\phi(p_2, \xi_W)\Delta_\phi(p_1, \xi_W)$, nor a term of the form $\Delta_\chi(q, \xi_Z)\Delta_\phi(p_2, \xi_W)$. Therefore, for the final cancellation to occur, the cofactors in front of all such linearly independent terms must individually vanish. This last condition gives rise to the advertised WI.

Let us first look at terms carrying only a gauge dependent factor of $\Delta_\chi(q,\xi_Z)$. Such terms can arise only from the diagrams shown in Fig.9. In what follows we use the WI of the boson self-energies Eqs.(5.4– 5.9) as well as the WI of the tree-level three vector boson vertex

$$q^\mu \, \Gamma_{\mu\alpha\beta}^{VW^-W^+}(q,p_1,p_2) = g_V \left[ U_W^{-1}(p_2)_{\alpha\beta} - U_W^{-1}(p_1)_{\alpha\beta} \right], \tag{5.10}$$

and pull out the common factor $(\bar{e}Ze)_\nu \, (\bar{\nu}W^+e)_\rho \, (\bar{e}W^-\nu)_\sigma$. Then, the $\Delta_\chi(q,\xi_Z)$ gauge dependent part is given by:

$$\frac{1}{M_Z^2}\Delta_\chi(q,\xi_Z) \, \{ \, C_v \; + \; C_{se} \; + \; C_m^P \, \}^{\nu\rho\sigma} \tag{5.11}$$

where $C_v, C_{se}, C_m^P$ are the contributions of the vertex, self-energy, and the pinched monodromic graphs, respectively:

$$C_v^{\nu\rho\sigma} = U_W^{\alpha\rho}(1) \, U_W^{\beta\sigma}(2) \, q^\nu \left[ q^\mu \widehat{\Gamma}_{\mu\alpha\beta}^W + iM_W \widehat{\Gamma}_{\alpha\beta}^\zeta \right] \tag{5.12}$$

$$C_{se}^{\nu\rho\sigma} = -U_W^{\alpha\rho}(1) \, U_W^{\beta\sigma}(2) \, q^\nu \, gc \left( \widehat{\Pi}_{\alpha\beta}^W(1) - \widehat{\Pi}_{\alpha\beta}^W(2) \right) \; - gcH^{\nu\rho\sigma} \tag{5.13}$$

$$(C_m^P)^{\nu\rho\sigma} = gcH^{\nu\rho\sigma} \tag{5.14}$$

with

$$\begin{aligned}
H^{\nu\rho\sigma} \;=\; & q^\nu \left[ U_W^{\alpha\rho}(1)\widehat{\Pi}_{\alpha\beta}^W(1) \, U_W^{\beta\sigma}(1) \;-\; U_W^{\alpha\rho}(2) \, \widehat{\Pi}_{\alpha\beta}^W(2) \, U_W^{\beta\sigma}(2) \right] \\
& + q^\alpha \, \widehat{\Pi}_{\alpha\beta}^Z(q) \, U_Z^{\beta\nu}(q) \, [ \, U_W^{\sigma\rho}(1) \;-\; U_W^{\sigma\rho}(2) \, ]
\end{aligned} \tag{5.15}$$

where 1, 2 in the arguments means $p_1$ and $p_2$, respectively.

Since the gauge independence of the amplitude requires that the sum $C_v \; + \; C_{se} \; + \; C_m^P$ in Eq.(5.11) must vanish, we arrive at the following WI, relating the $ZWW$ and $\chi WW$ vertices:

$$q^\mu \widehat{\Gamma}_{\mu\alpha\beta}^{ZW^-W^+} + iM_Z \widehat{\Gamma}_{\alpha\beta}^{\chi W^-W^+} = gc \left[ \widehat{\Pi}_{\alpha\beta}^W(1) - \widehat{\Pi}_{\alpha\beta}^W(2) \right] \tag{5.16}$$

Repeating similar steps and requiring the cancellation of the $\Delta_\phi(1)$ and $\Delta_\phi(2)$ gauge dependencies, we obtain the following Ward identities, respectively

$$p_1^\alpha \widehat{\Gamma}_{\mu\alpha\beta}^{ZW^-W^+} + iM_W \widehat{\Gamma}_{\mu\beta}^{Z\phi^-W^+} = gc \left[ \widehat{\Pi}_{\mu\beta}^W(2) - \widehat{\Pi}_{\mu\beta}^Z(q) - \frac{s}{c}\widehat{\Pi}_{\mu\beta}^{Z\gamma}(q) \right] \tag{5.17}$$

$$p_2^\beta \widehat{\Gamma}^{ZW^-W^+}_{\mu\alpha\beta} + iM_W \widehat{\Gamma}^{Z\phi^-W^+}_{\mu\alpha} = gc \left[ \widehat{\Pi}^Z_{\mu\beta}(q) + \frac{s}{c}\widehat{\Pi}^{Z\gamma}_{\mu\beta}(q) - \widehat{\Pi}^W_{\mu\beta}(1) \right] \quad (5.18)$$

The WI for the $\widehat{\Gamma}^{\gamma W^-W^+}$ vertex can be derived in a similar manner. We have:

$$p_1^\alpha \widehat{\Gamma}^{\gamma W^-W^+}_{\mu\alpha\beta} + iM_W \widehat{\Gamma}^{\gamma\phi^-W^+}_{\mu\beta} = gs \left[ \widehat{\Pi}^W_{\mu\beta}(2) - \widehat{\Pi}^\gamma_{\mu\beta}(q) - \frac{c}{s}\widehat{\Pi}^{\gamma Z}_{\mu\beta}(q) \right] \quad (5.19)$$

$$p_2^\beta \widehat{\Gamma}^{ZW^-W^+}_{\mu\alpha\beta} + iM_W \widehat{\Gamma}^{Z\phi^-W^+}_{\mu\alpha} = gs \left[ \widehat{\Pi}^\gamma_{\mu\beta}(q) + \frac{c}{s}\widehat{\Pi}^{\gamma Z}_{\mu\beta}(q) - \widehat{\Pi}^W_{\mu\beta}(1) \right] \quad (5.20)$$

which are the counterparts of Eq.(5.17) and Eq.(5.18). It is elementary to derive additional WI, through straightforward algebraic manipulations of the WI listed above. For example, the WI

$$iM_Z p_1^\alpha \widehat{\Gamma}^{\chi W^-W^+}_{\alpha\beta} - iM_W q^\alpha \widehat{\Gamma}^{Z\phi^-W^+}_{\alpha\beta} = gc \left[ p_1^\alpha \widehat{\Pi}^W_{\alpha\beta}(1) + p_2^\alpha \widehat{\Pi}^W_{\alpha\beta}(2) + q^\alpha \widehat{\Pi}^Z_{\alpha\beta}(q) \right] \quad (5.21)$$

or equivalently

$$p_1^\alpha \widehat{\Gamma}^{\chi W^-W^+}_{\alpha\beta} - c\, q^\alpha \widehat{\Gamma}^{Z\phi^-W^+}_{\alpha\beta} = gc \left[ c\widehat{\Pi}^+_\beta(1) + c\widehat{\Pi}^+_{\alpha\beta}(2) + \widehat{\Pi}^{Z\chi}_\beta(q) \right] \quad (5.22)$$

can be immediately obtained from Eq.(5.16) and Eq.(5.17), after contracting them with the appropriate four-momenta, and using the WIs of the self energies and the fact that $\widehat{\Pi}^{Z\gamma}_{\alpha\beta}(q)$ is transverse.

Finally, WIs where the g.i. TBV are contracted with two or three momenta can be easily derived, by demanding the cancellation of gauge dependencies stemming from terms with more than one Goldstone boson propagator.

It is interesting to notice that an equation analogous to Eq.(5.16) for the $\widehat{\Gamma}^{\gamma W^-W^+}$ vertex, cannot be derived via this procedure. The reason is simply that all residual dependence on $\xi_\gamma$ automatically disappears from the final expressions, due to current conservation, e.g. $q^\mu J^\gamma_\mu = 0$. In order to derive the remaining WI one must choose a gauge–fixing procedure like the axial or light–cone gauge, where the dependence of the gauge–boson legs on the gauge parameter does not vanish due to current conservation. In fact, this was the way the PT was originally implemented by Cornwall, when constructing the one–loop g.i. gluon self-energy [2]. In the axial (light–cone) gauge, for example, the tree–level propagator for the photon reads

$$\Delta^{\mu\nu}_\gamma(q,n) = \frac{1}{q^2}\left[ g_{\mu\nu} - \frac{n_\mu q_\nu + n_\nu q_\mu}{n \cdot q} \right], \quad (5.23)$$

where $n_\mu$ is the gauge–fixing parameter (in the light–cone gauge $n_\mu n^\mu = 0$). So, after using current conservation, the $n$ with the appropriate Lorentz index will vanish, but the other $n$ will survive, and will only vanish if the desired WI are satisfied. Finally we obtain :

$$\begin{aligned} q^{\mu}\widehat{\Pi}_{\mu\nu}^{\gamma\gamma}(q) &= 0 \\ q^{\mu}\widehat{\Pi}_{\mu\nu}^{\gamma Z}(q) &= 0 \\ q^{\mu}\widehat{\Pi}_{\mu}^{\gamma\chi}(q) &= 0 \quad \Rightarrow \quad \widehat{\Pi}_{\mu}^{\gamma\chi}(q) = 0 \end{aligned} \qquad (5.24)$$

and

$$q^{\mu}\widehat{\Gamma}_{\mu\alpha\beta}^{\gamma W^-W^+} = gs \left[\widehat{\Pi}_{\alpha\beta}^W(1) - \widehat{\Pi}_{\alpha\beta}^W(2)\right], \qquad (5.25)$$

which was first proved in [23] by an explicit one-loop calculation. Clearly, similar WI can be derived for the gluon self–energy and three gluon vertex in QCD.

All previous WI are the one–loop generalizations of the respective *tree level* WI. As shown in this section, their validity is crucial for the gauge independence of the S–matrix. It is important to emphasize that these WI make no reference to ghost terms, unlike the corresponding Slavnov-Taylor identities satisfied by the conventional, gauge–dependent vertices.

The WI derived in this section are also true in the context of the BFM. In fact, in the BFM framework they are true to all orders in perturbation theory; their validity is enforced by the requirement that the Lagrangian is invariant under gauge transformations of the *background fields*. It should be emphasized however that the Green's functions of the background fields, which satisfy the aforementioned WIs, display in general a residual dependence on the parameter $\xi_Q$ used to gauge fix the quantum gauge fields. As shown in [27] this remaining gauge dependence can be eliminated by the straightforward application of the PT in the context of the BFM. The analysis presented in this section indicates that these "naive", WIs are not an exclusive property of the BFM, but can be recovered for any type of gauge fixing procedure via the PT algorithm. Strictly speaking the WIs we have presented are valid to one loop order. This is so because our derivation relies on the ability to construct the $\xi$–independent Green's functions (shown as blobs in Fig.8) with the PT algorithm, which has only been tested at one–loop. If one assumes that this procedure of isolating $\xi$–independent blobs can be generalized to higher orders in perturbation theory, the generalization of the WI to higher orders will be relatively straightforward. Even though such an assumption is rather plausible no such proof exists.

# 6 Proof of the Ward Identities

## 6.1 Feynman gauge

In this section we prove by an explicit calculation the first of the Ward identities derived in the previous section, namely $q^\mu \widehat{\Gamma}^{ZW^-W^+}_{\mu\alpha\beta} + iM_Z \widehat{\Gamma}^{\chi W^-W^+}_{\alpha\beta} = gc\left[\widehat{\Pi}^W_{\alpha\beta}(1) - \widehat{\Pi}^W_{\alpha\beta}(2)\right]$. We work in the Feynman gauge, where $\xi_i = 1$ for $i = \gamma, W, Z$. To that end, it is more economical to act with $q^\mu$ directly on the individual graphs of $\widehat{\Gamma}^Z_{\mu\alpha\beta}$ and try to generate the r.h.s. of Eq.(5.16). The Feynman diagrams contributing to the g.i. $W$ self-energies of the r.h.s. are shown in Fig.10. The closed expression for the g.i. $W$ self-energy has been obtained in [16] and it is

$$\widehat{\Pi}^W_{\alpha\beta}(q) = \Pi^W_{\alpha\beta}(q)|_{\xi=1} + 4U^{-1}_{\alpha\beta}(q)\left[s^2 I_{W\gamma}(q) + c^2 I_{WZ}(q)\right] \tag{6.1}$$

We also emphasize that all necessary cancellations between graphs or parts of graphs are evident before any of the loop momentum integrations are carried out.

To begin with, we notice that, all pinch parts originating from the top (neutral current) fermion line, automatically cancel in the l.h.s. Eq.(5.16), by virtue of the second and third of Eqs.(3.13).

We start by considering the fermion graphs Fig.5 (1) and Fig.6 (1). This subset of graphs is automatically g.i., and receives therefore no pinch contributions. After a straightforward calculation we obtain ( in what follows we have pulled out a common factor of $gc$ from the r.h.s. of all equations ) :

$$q^\mu V^1_{\mu\alpha\beta} + iM_Z \mathcal{S}^1_{\alpha\beta} = \Pi^1_{\alpha\beta}(1) - \Pi^1_{\alpha\beta}(2) \tag{6.2}$$

where $\Pi^1_{\alpha\beta}$ corresponds to the self-energy diagram (1) of Fig.10.

The remaining diagrams can be divided into three classes, depending on the type of internal boson propagators they contain. Following the notation of Eq.(4.3), these classes are denoted as (i) $WWV$ diagrams, where $V = \gamma, Z$ ( Fig.5 (2)–(22), Fig.6 (2)–(13) ) (ii) $WWH$ diagrams ( Fig.5 (23)–(28), Fig.6 (14),(15) ), and (iii) $ZHW$ diagrams ( Fig.5 (29)–(34), Fig.6 (16)–(21) ).

**WWV graphs :**

Vector boson graphs

$$\sum_{i=2}^{8} q^\mu V^i_{\mu\alpha\beta} = \sum_{i=2}^{3}\left[\Pi^i_{\alpha\beta}(1) - \Pi^i_{\alpha\beta}(2)\right]$$
$$- \sum_{V=\gamma Z} g_V^2 \Delta M^2_{WV} \int (WWV)\; q^\rho \Gamma_{\rho\alpha\beta}(q, p_1-k, p_2+k)$$

$$+ \sum_{V=\gamma Z} g_V^2 \int W(k+p_1)V(k)\, k_\alpha q_\beta + (1 \leftrightarrow 2)$$

$$- \sum_{V=\gamma Z} g_V^2 \int (WWV)\, q\cdot(k-p_2)\, k_\alpha(k-p_2)_\beta + (1 \leftrightarrow 2)$$

$$-2g^2 \left[ U_W^{-1}(1)_{\alpha\beta} - U_W^{-1}(2)_{\alpha\beta}\right] \int W(k+p_1)W(k-p_2)$$

$$- U_W^{-1}(1)_{\alpha\rho} \sum_{V=\gamma Z} g_V^2 \int (WWV)\left[q^\rho k_\beta + q^\lambda \Gamma_{\lambda\rho\beta}(k-p_2,-k,p_2)\right] + (1\leftrightarrow 2) \quad (6.3)$$

where the tree-level WI of Eq.(5.10), as well as the identities given in equations (4.11) and (4.12) of [23], have been used. The notation $(1 \leftrightarrow 2)$ means to interchange in the precceding term $p_1 \leftrightarrow -p_2$ and $\alpha \leftrightarrow \beta$. From the terms appearing in the r.h.s. of Eq.(6.3) only the first is part of the r.h.s. of the WI we attempt to prove. All other terms will cancel against other contributions from the remaining graphs. In particular, the left-over term of the second line will cancel against similar terms coming from the graphs which contain unphysical Goldstone bosons. Similarly, the terms in the next two lines of Eq.(6.3) will cancel against corresponding left-overs from the ghost graphs. Finally, the last two lines of Eq.(6.3), which display the characteristic pinch structure, will cancel against some of the pinch contributions to the $ZWW$ vertex. All these cancellations will become evident in what follows.

We next consider the ghost graphs (Fig.5 (9)–(12) and Fig.6 (2)–(5)). We have:

$$\sum_{i=9}^{12} q^\mu V^i_{\mu\alpha\beta} = \sum_{i=4}^{7}\left[\Pi^i_{\alpha\beta}(1) - \Pi^i_{\alpha\beta}(2)\right]$$

$$- \sum_{V=\gamma Z} g_V^2 \int W(k+p_1)V(k)\, k_\alpha q_\beta + (1\leftrightarrow 2)$$

$$+ \sum_{V=\gamma Z} g_V^2 \int (WWV)\, q\cdot(k-p_2)\, k_\alpha(k-p_2)_\beta + (1\leftrightarrow 2) \quad (6.4)$$

and

$$iM_Z \sum_{i=2}^{5} \mathcal{S}^i_{\alpha\beta} = -\frac{M_W^2}{2c^2} \sum_{V=\gamma Z} g_V^2 \int (WWV)\, (k_\alpha p_{2\beta} + k_\beta p_{1\alpha}) \quad (6.5)$$

We see that the left-over terms of Eq.(6.4) cancel against parts of Eq.(6.3) as advertised. The contribution of Eq.(6.5) will cancel against pinch contributions to the $\chi WW$ vertex. Similarly, the graphs containing unphysical Goldstone bosons (Fig.5 (13)–(22) and Fig.6 (6)–(13)) yield:

$$\sum_{i=13}^{16} q^\mu V^i_{\mu\alpha\beta} + iM_Z \sum_{i=6}^{9} \mathcal{S}^i_{\alpha\beta} = g^2 s^2 M_W^2 \sum_{V=\gamma,Z} b_V \int (WWV)\, q^\rho \Gamma_{\rho\alpha\beta}(q, p_1-k, p_2+k)$$

$$+ g^2 M_W^2 \frac{s^2}{2c^2}\left[U_W^{-1}(1)_{\alpha\beta} - U_W^{-1}(2)_{\alpha\beta}\right] \sum_{V=\gamma,Z} b_V J_{WWV} \quad (6.6)$$

$$\sum_{i=17}^{18} q^\mu V^i_{\mu\alpha\beta} + iM_Z \sum_{10}^{13} \mathcal{S}^i_{\alpha\beta} = \sum_{i=8}^{9} \left[ \Pi^i_{\alpha\beta}(1) - \Pi^i_{\alpha\beta}(2) \right] \tag{6.7}$$

$$\sum_{i=19}^{21} q^\mu V^i_{\mu\alpha\beta} = \Pi^{10}_{\alpha\beta}(1) - \Pi^{10}_{\alpha\beta}(2)$$
$$- \frac{g^2}{8c^2} \int W(k) \left[ Z(k+p_1) - Z(k-p_2) \right] (2k+p_1)_\alpha (2k-p_2)_\beta \tag{6.8}$$

$$q^\mu V^{22}_{\mu\alpha\beta} = 0 \tag{6.9}$$

The first term in the r.h.s. of Eq.(6.6) cancels the appropriate term in Eq.(6.3), after employing the elementary identity $-gcg_V^2 \Delta M_{WV}^2 = -b_V g^3 cs^2 M_W^2$, where $b_\gamma = +1$ and $b_Z = -1$. The second term in the r.h.s. of Eq.(6.6) will cancel against pinch contributions to the $\chi WW$ vertex. Finally, we note that the left-over term in the r.h.s. of Eq.(6.8) contains only two internal propagators.

The pinch parts give:

$$q^\mu \left[ B^-_{\mu\rho\beta} U_W^{-1}(p_1)^\rho_\alpha + B^+_{\mu\alpha\rho} U_W^{-1}(p_2)^\rho_\beta \right] = \tag{6.10}$$
$$2 \left[ U_W^{-1}(p_1)_{\alpha\beta} + U_W^{-1}(p_2)_{\alpha\beta} \right] \sum_{V=\gamma,Z} g_V^2 \left[ I_{WV}(p_1) - I_{WV}(p_2) \right]$$
$$+ U_W^{-1}(p_2)_{\beta\rho} \sum_{V=\gamma,Z} g_V^2 \int (WWV) \left[ q^\rho k_\alpha + q^\lambda \Gamma_{\lambda\alpha\rho}(-k-p_1, p_1, k) \right] + (1 \leftrightarrow 2)$$

$$-2g^2 q^\mu \Gamma_{\mu\alpha\beta} [\ldots] = 2g^2 \left[ U_W^{-1}(p_1)_{\alpha\beta} - U_W^{-1}(p_2)_{\alpha\beta} \right] [\ldots] \tag{6.11}$$

where the ellipses in the square brackets in Eq.(6.11) represent the terms of the second line of Eq.(4.34) multiplying $\Gamma_{\mu\alpha\beta}$.

$$\sum_{i=1}^{4} q^\mu \left( R^i_{\mu\alpha\beta} + L^i_{\mu\alpha\beta} \right) = g^2 M_W^2 [p_{2\beta} q_\alpha - p_{1\alpha} q_\beta] \left[ \frac{s^2}{c^2} J_{WW\gamma} + \frac{(1-2s^2)}{2c^2} J_{WWZ} \right] \tag{6.12}$$

$$iM_Z \sum_{i=1}^{2} \left( \mathcal{R}^i_{\alpha\beta} + \mathcal{L}^i_{\alpha\beta} \right) = g^2 M_W^2 \frac{s^2}{2c^2} \left[ U^{-1}_{\alpha\beta}(p_1) - U^{-1}_{\alpha\beta}(p_2) \right] \sum_{V=\gamma Z} b_V J_{WWV} \tag{6.13}$$

$$iM_Z \left( \mathcal{R}^3_{\alpha\beta} + \mathcal{L}^3_{\alpha\beta} \right) = g^2 M_W^2 \frac{s^2}{2c^2} \int (WW\gamma) \left[ (k-2q)_\alpha p_{2\beta} + p_{1\alpha}(k+2q)_\beta \right] \tag{6.14}$$

$$iM_Z \sum_{i=4}^{5} \left( \mathcal{R}_{\alpha\beta}^i + \mathcal{L}_{\alpha\beta}^i \right) = + g^2 \frac{M_W^2}{2} \int (WWZ) \left[ k_\alpha p_{2\beta} + p_{1\alpha} k_\beta \right]$$
$$- g^2 M_W^2 \frac{(1-2s^2)}{2c^2} \left[ p_{2\beta} q_\alpha - p_{1\alpha} q_\beta \right] J_{WWZ} \quad (6.15)$$

At this point we notice that the contributions of the pinch parts cancel all the remaining left-overs of all other graphs we have considered thus far, except for the left-over term of Eq.(6.8). If we now collect all $W$ self-energy terms in the r.h.s., we notice that all pinch and regular graphs have already appeared, except for the two graphs containing an internal Higgs boson, shown in Fig.10 (11) and (12)

Next we consider the $WWH$ graphs

$$\sum_{i=23}^{25} q^\mu V_{\mu\alpha\beta}^i + iM_Z \sum_{i=14}^{15} \mathcal{S}_{\alpha\beta}^i = \Pi_{\alpha\beta}^{11}(1) - \Pi_{\alpha\beta}^{11}(2)$$
$$- \frac{g^2 M_W^2}{4c^2} \int (WWH) \left[ k_\alpha p_{2\beta} + k_\beta p_{1\alpha} \right]$$
$$- \frac{g^2 M_W^2}{2} [p_{1\alpha} p_{1\beta} - p_{2\alpha} p_{2\beta}] J_{WWH} \quad (6.16)$$

$$\sum_{i=26}^{28} q^\mu V_{\mu\alpha\beta}^i = \Pi_{\alpha\beta}^{12}(1) - \Pi_{\alpha\beta}^{12}(2) \quad (6.17)$$
$$- \frac{g^2}{8c^2} \int W(k) \left[ H(k+p_1) - H(k-p_2) \right] (2k+p_1)_\alpha (2k-p_2)_\beta$$

The relevant pinch diagrams of this class contributing to the $ZWW$ vertex give

$$q^\mu \left( R_{\mu\alpha\beta}^5 + L_{\mu\alpha\beta}^5 \right) = \frac{g^2 M_W^2}{2} [p_{1\alpha} p_{1\beta} - p_{2\alpha} p_{2\beta}] J_{WWH} \quad (6.18)$$

whereas the ones contributing to the $\chi WW$ vertex give

$$iM_Z \left( \mathcal{R}_{\alpha\beta}^6 + \mathcal{L}_{\alpha\beta}^6 \right) = \frac{g^2 M_W^2}{4c^2} \int (WWH) \left[ k_\alpha p_{2\beta} + k_\beta p_{1\alpha} \right] \quad (6.19)$$

It is now evident that all the $W$ self-energy terms which constitute the r.h.s. of the WI have already appeared. The only two redundant terms are (i) the left-over term of Eq.(6.8), and (ii) the left-over term of Eq.(6.18), which survives after Eq.(6.16)–Eq.(6.19) have been added by parts. Like the term in (i) it also contains two internal propagators. Both terms will cancel exactly against the entire contribution of the graphs belonging to the $ZHW$ class, which we now proceed to evaluate.

We will only evaluate the diagrams where the Higgs boson appears on the left. The mirror graphs, with the Higgs boson on the right, can be treated in an exactly analogous way.

$$q^\mu V^{30}_{\mu\alpha\beta} = -\frac{g^2 M_W^2}{c^2} \int (HZW) \, q^\mu \Gamma_{\mu\alpha\beta}(-k-p_1, p_1, k) \tag{6.20}$$

$$\begin{aligned}
iM_Z \mathcal{S}^{17}_{\alpha\beta} &= \frac{g^2 M_W^2}{c^2} \int (HZW) \, q^\mu \Gamma_{\mu\alpha\beta}(-k-p_1, p_1, k) - g^2 \frac{M_W^2}{2c^2} U_W^{-1}(p_2)_{\alpha\beta} J_{HZW} \\
&\quad - g^2 \frac{M_W^2}{2c^2} \int (HZW) \, k_\alpha k_\beta + g^2 \frac{M_W^2}{2c^2} g_{\alpha\beta} I_{HZ}(q)
\end{aligned} \tag{6.21}$$

The last term in the r.h.s. of the last equation will cancel against an equal and opposite contribution coming from the mirror diagram $S^{18}_{\alpha\beta}$

$$q^\mu V^{32}_{\mu\alpha\beta} = -\frac{g^2 M_Z^2 s^2}{2c^2} q_\beta \int (HZW)(2k+p_1)_\alpha \tag{6.22}$$

$$iM_Z \mathcal{S}^{19}_{\alpha\beta} = \frac{g^2 M_Z^2 s^2}{2c^2} q_\beta \int (HZW)(2k+p_1)_\alpha - g^2 \frac{M_Z^2 s^2}{4c^2} \int (HZW) \, (2k+p_1)_\alpha (k-p_2)_\beta \tag{6.23}$$

$$\begin{aligned}
q^\mu V^{34}_{\mu\alpha\beta} &= \frac{g^2}{8c^2} \left( M_Z^2 - M_H^2 \right) \int (HZW) \, (2k+p_1)_\alpha (2k-p_2)_\beta \\
&\quad + \frac{g^2}{8c^2} \int W(k) \left[ H(k+p_1) - Z(k-p_2) \right] (2k+p_1)_\alpha (2k-p_2)_\beta
\end{aligned} \tag{6.24}$$

$$iM_Z \mathcal{S}^{21}_{\alpha\beta} = \frac{g^2}{8c^2} M_H^2 \int (HZW) \, (2k+p_1)_\alpha (2k-p_2)_\beta \tag{6.25}$$

The pinch parts are :

$$q^\mu L^6_{\mu\alpha\beta} + iM_Z \mathcal{L}^7_{\alpha\beta} = -g^2 \frac{M_W^2}{2c^2} p_{1\alpha} \int (HZW) \, k_\beta \tag{6.26}$$

$$iM_Z \left( \mathcal{R}^8_{\alpha\beta} + \mathcal{R}^{8cr}_{\alpha\beta} \right) = g^2 \frac{M_W^2}{2c^2} U_W^{-1}(p_2)_{\alpha\beta} J_{HZW} \tag{6.27}$$

$$iM_Z \left( \mathcal{R}^9_{\alpha\beta} + \mathcal{R}^{9cr}_{\alpha\beta} \right) = g^2 \frac{M_Z^2(1-2s^2)}{8c^2} p_{2\beta} \int (HZW)(2k+p_1)_\alpha \tag{6.28}$$

When all the above equations, together with the corresponding contributions from the mirror graphs, are added by parts, all terms in the r.h.s. cancel among each other as expected, except for the terms with two internal propagators, from Eq.(6.24) and the mirror graph result, which exactly cancel the left-over terms mentioned previously, (i) and (ii). This concludes the proof of the advertised WI of Eq.(5.16), which is a central result of this paper. It is obvious from the previous proof that the pinch parts are instrumental for the validity of Eq.(5.16).

## 6.2 Unitary gauge

The fact that the WIs of Green's functions constructed via the PT hold regardless of the gauge in which one chooses to work, can be most effectively demonstrated by proving their validity in different gauges. Although the usual graphs of a Green's function as well as its pinch parts assume different forms in different gauges, when summed they nevertheless combine into a unique expression independent of any specific gauge. In this section we will work in the unitary gauge, where additional pinch parts can originate from the longitudinal parts of the gauge boson propagators.

We note that, even though the unitary gauge has been traditionally considered pathological, in the sense that it gives rise to non–renormalizable Green's functions , in the context of the PT it can be treated on an equal footing as the renormalizable $R_\xi$ gauges. In particular, as shown in [29] the application of the PT in the context of the unitary gauge gives rise to *renormalizable* Green's functions which are in fact identical to the $\xi$–independent Green's functions obtained in the framework of the $R_\xi$ gauges.

Applying the PT to the case of the three boson vertices in the unitary gauge, we have verified that the WI of Eq.(5.16) and Eq.(5.25) again hold true. We point out, that although the usual vertex graphs are fewer in this gauge, the graphs which can contribute pinch parts are quite numerous, a fact that makes calculations lengthier. We therefore do not present the entire proof of the WI, but only outline the steps in its derivation.

The usual $\gamma W^- W^+$ vertex diagrams in this gauge are shown in Fig.5 (1)–(8) and (23), while for the $ZW^- W^+$ vertex we have the additional graphs (29) and (30) of the same figure. The relevant $W$ self energy diagrams in the unitary gauge are those shown in Fig.10 (1)–(3) and (11). The vertex graphs will be denoted as $\mathcal{V}^i_{\mu\alpha\beta}$, and the self energy graphs as $\mathcal{U}^i_{\alpha\beta}$, where the index $i$ counts the corresponding graphs of Fig.5 and Fig.10.

The interesting feature of the unitary gauge is that the WI of the $\gamma W^- W^+$ vertex is satisfied *separately* by the usual and pinch parts, as one can verify immediately.

For the fermion graphs Eq.(6.2) holds as usual since they are gauge invariant, e.g. $V^1_{\mu\alpha\beta} = \mathcal{V}^1_{\mu\alpha\beta}$ and $\Pi^1_{\alpha\beta} = \mathcal{U}^1_{\alpha\beta}$. The boson graphs give :

$$q^\mu \mathcal{V}^{2,3}_{\mu\alpha\beta} + q^\mu \mathcal{V}^{6,5}_{\mu\alpha\beta} + q^\mu \mathcal{V}^{8,7}_{\mu\alpha\beta} = gc \left[ \mathcal{U}^{2,3}_{\alpha\beta}(1) - \mathcal{U}^{2,3}_{\alpha\beta}(2) \right] \tag{6.1}$$

and

$$q^\mu \mathcal{V}^4_{\mu\alpha\beta} = 0 \tag{6.2}$$

From the Higgs diagram we get :

$$q^\mu \mathcal{V}^{23}_{\mu\alpha\beta} = gc \left[ \mathcal{U}^{11}_{\alpha\beta}(1) - \mathcal{U}^{11}_{\alpha\beta}(2) \right] \tag{6.3}$$

For the $\gamma WW$ vertex the above equations when summed give :

$$q^\mu \mathcal{V}^{\gamma}_{\mu\alpha\beta} = gc \left[ \mathcal{U}^{W}_{\alpha\beta}(1) - \mathcal{U}^{W}_{\alpha\beta}(2) \right] \tag{6.4}$$

Evidently the Ward identity holds already for the usual vertex graphs, *before* any pinch contributions are included. One can then verify that the vertex like pinch contributions in the unitary gauge $\left(\mathcal{V}^{\gamma}_{\mu\alpha\beta}\right)^P$ (some of the additional ones, specific to the unitary gauge, are shown in Fig.11) , when contracted with $q^\mu$, yield :

$$q^\mu \left(\mathcal{V}^{\gamma}_{\mu\alpha\beta}\right)^P = gc \left[ \mathcal{P}^{W}_{\alpha\beta}(1) - \mathcal{P}^{W}_{\alpha\beta}(2) \right] \tag{6.5}$$

where $\mathcal{P}^{W}_{\alpha\beta}$ are the relevant propagator like parts, to be appended to the $W$ self-energy in the unitary gauge. As shown in [29] the $W$ self-energy obtained via the PT in the unitary gauge is identical to the one obtained via the PT in the context of the $R_\xi$ gauges, namely :

$$\widehat{\Pi}^{W}_{\alpha\beta}(q) = \mathcal{U}^{W}_{\alpha\beta} + \mathcal{P}^{W}_{\alpha\beta} \ . \tag{6.6}$$

Adding Eqs.(6.4) and (6.5) by parts we arrive again at the advertised WI of Eq.(5.25).

For the $ZWW$ vertex the proof proceeds in an analogous way. For the class of graphs that are common to both vertices $\gamma WW$ and $ZWW$ the proof is identical. The WI is again satisfied *separately* by the regular vertex graphs and the pinch graphs. There are however two additional classes of contributions that need be considered. First, there are the extra regular vertex graphs 29 and 30 of Fig.5 along with similar box graphs that will contribute a pinch part to the vertex ; all the above graphs contain a Higgs particle. Second, unlike the photon case, the vertex like pinch parts originating from boxes where the pinching takes place at the leg of the $Z$, (fermion line on the top) do not vanish when contracted with $q^\mu$. Since all propagator graphs of the $W$ have already appeared on the left hand side of the WI, the sole role of these graphs is to provide a left over expression which is recognized as being equal to $\widehat{\Gamma}^{\chi W^- W^+}_{\alpha\beta}$. Of course, in the context of the unitary gauge this expression cannot be identified with a $\chi WW$ vertex, because there are no $\chi$-fields to begin with.

## 6.3   Feynman background field gauge

In this section we prove the validity of the WI of Eqs.5.16 and Eq.(5.25) in the Feynman gauge of the BFM.

In the BFM every bosonic field is decomposed into two parts, the quantum field $\Phi$ and the background field $\widehat{\Phi}$, e.g. $\Phi \to \Phi + \widehat{\Phi}$. In the path integral one integrates the quantum fields only, whereas the background fields are treated as additional sources ; consequently only the quantum fields appear inside loops.

The ordinary gauge transformation of the gauge fields, for example $W_\mu^a$, $a = 1, 2, 3$ and $B_\mu$ in the case of an $SU(2) \times U(1)$ group, is also split into two transformations. One of them corresponds to an ordinary gauge transformation, but only for the background fields $\widehat{W}_\mu^a, \widehat{B}_\mu$, and is therefore called a background gauge transformation. By judiciously adding a non conventional gauge fixing term for the quantum fields we can promote this transformation to a symmetry of the Lagrangian. Therefore the Green's functions of the background fields are guaranteed to be background gauge invariant, namely $\Gamma(\hat{W}_\mu^a \hat{B}_\nu ...) = \Gamma(\hat{W}_\mu'^a \hat{B}_\nu' ...)$. As a result of this invariance the naive WI of section 5 are satisfied. Notice however that these Green's functions depend in general on the gauge parameters $\xi_W, \xi_B$, used to gauge fix the quantum fields $W_\mu^a$ and $B_\mu$ which appear inside their loops. In this formulation, S-matrix elements are calculated by forming trees of background Green's functions connected to each other by tree level background field propagators ; at this point the background fields also require gauge fixing.

This gauge fixing is completely independent from the gauge fixing of the quantum fields, and the parameters $\hat{\xi}_W$, $\hat{\xi}_B$ may be in general different from the parameters $\xi_W$, $\xi_B$.

We choose to work in the Feynman gauge of the BFM where $\xi_W = \xi_B \equiv \xi_Q = 1$. As was shown in [25] [27], at the one loop level, this particular gauge choice gives rise to background Green's functions which are identical to the g.i. Green's functions constructed via the PT. No formal understanding of this correspondence has yet been established ; the aforementioned agreement has been verified by comparing all Green's functions constructed so far at one loop via the PT with the corresponding BFM Green's functions . The operational reason for this identity of results is that pinching turns out to be zero in this particular gauge . To this end we remind the reader that pinch parts can originate in three ways :
i) from the longitudinal part of gauge boson propagators,
ii) from three gauge-boson vertices, and
iii) from vertices with two Goldstone bosons and one gauge boson.
All these can provide the appropriate momenta which when contracted with a $\gamma$-matrix will cancel a fermion propagator. By simple inspection of the Feynman rules of this gauge one immediately recognizes that all the necessary pieces that could generate pinch terms are missing. First of all, since this is a Feynman type of gauge there are no longitudinal parts for the gauge boson propagators. Secondly, one observes that in this gauge the three gauge boson vertex between a background and two quantum gauge fields ( which is gauge dependent even at tree level ) assumes the form

$$\Gamma_{\alpha\beta\gamma}(q,k,-q-k) = \Gamma^F_{\alpha\beta\gamma} + (1 - \frac{1}{\xi_Q})\Gamma^P_{\alpha\beta\gamma} \qquad (6.1)$$

where

$$\Gamma^F_{\alpha\beta\gamma} = (2k+q)_\alpha g_{\beta\gamma} - 2q_\beta g_{\gamma\alpha} + 2q_\gamma g_{\alpha\beta} \qquad (6.2)$$

and

$$\Gamma^P_{\alpha\beta\gamma} = -k_\beta g_{\gamma\alpha} - (k+q)_\gamma g_{\alpha\beta}. \qquad (6.3)$$

We see immediately that by setting $\xi_Q = 1$, the $\Gamma^P_{\alpha\beta\gamma}$ part, which is the only one that can pinch, disappears. Finally, the elementary vertices of the form $\hat{\phi}\phi G$ where $\hat{\phi},\phi$ are scalars (Higgs or unphysical would be Goldstone bosons) and $G_\mu$ a quantum gauge field $(W^a_\mu, B_\mu)$, depend only on the momentum carried by the background field $\hat{\phi}$, namely

$$\Gamma^{\hat{\phi}\phi G}_\mu(q,k,-q-k) \propto q_\mu \qquad (6.4)$$

Therefore they also cannot trigger pinching. Consequently, since pinching has been rendered trivial (zero) in this gauge, one readily concludes that the Green's functions constructed via the PT in any gauge will be identical to the conventional Green's functions of the Feynman gauge of the BFM, i.e.

$$\widehat{\Pi}^W_{\alpha\beta} = \Pi^{\hat{W}}_{\alpha\beta} \quad , \quad \widehat{\Gamma}^{ZW^-W^+}_{\mu\alpha\beta} = \Gamma^{\hat{Z}\hat{W}^-\hat{W}^+} \quad , \quad etc... \qquad (6.5)$$

We now proceed to the proof of the WIs of Eq.(5.16) and Eq.(5.25). We need to consider only the usual vertex and self energy graphs in this gauge. The vertex graphs are those of Fig.5 and Fig.6 plus the additional ones of Fig.11 and Fig.12. Note that for the $\chi W^-W^+$ vertex there are no ghost graphs in this gauge ; so for this paragraph the graphs of Fig.6 (2,3,4,5) are replaced by those of Fig.12 (2,3,4,5). The modifications needed for the W self energy graphs are that an additional graph (Fig.10 (13)) must be included and the pinch graph must be removed. In all these figures the external legs are now considered to be background fields.

We will use the same symbols for the various diagrams as in section 6.1, even though now, since we work in a different gauge, they correspond in general to different expressions. So $V^i_{\mu\alpha\beta}$ will correspond to a $ZW^-W^+$ vertex diagram , $S^i_{\alpha\beta}$ to a $\chi W^-W^+$ vertex and $\Pi^i_{\alpha\beta}$ to a $W$ self-energy graph.

By acting with $q^\mu$ on the three gauge boson vertex graphs we readily obtain the following results.

Fermion graphs :

$$q^\mu V^1_{\mu\alpha\beta} + iM_Z S^1_{\alpha\beta} = gc \left[ \Pi^1_{\alpha\beta}(1) - \Pi^1_{\alpha\beta}(2) \right] \qquad (6.6)$$

Gauge boson graphs :

$$q^\mu V^{2,3}_{\mu\alpha\beta} + q^\mu V^{4,5}_{\mu\alpha\beta} + q^\mu V^{6,7}_{\mu\alpha\beta} = gc \left[ \Pi^{2,3}_{\alpha\beta}(1) - \Pi^{2,3}_{\alpha\beta}(2) \right] \tag{6.7}$$

$$q^\mu V^{8}_{\mu\alpha\beta} = 0 \tag{6.8}$$

Ghost graphs :

$$q^\mu V^{9,10}_{\mu\alpha\beta} + q^\mu V^{35,36}_{\mu\alpha\beta} + q^\mu V^{37,38}_{\mu\alpha\beta} = gc \left[ \Pi^{5,4}_{\alpha\beta}(1) - \Pi^{4,5}_{\alpha\beta}(2) \right] \tag{6.9}$$

$$q^\mu V^{39}_{\mu\alpha\beta} = 0 \tag{6.10}$$

$$q^\mu V^{11,12}_{\mu\alpha\beta} + q^\mu V^{41,40}_{\mu\alpha\beta} + q^\mu V^{43,42}_{\mu\alpha\beta} = gc \left[ \Pi^{7,6}_{\alpha\beta}(1) - \Pi^{4,5}_{\alpha\beta}(2) \right] \tag{6.11}$$

$$q^\mu V^{44}_{\mu\alpha\beta} = 0 \tag{6.12}$$

In fact Eq.(6.9) is identical to Eq.(6.11) part by part and correspondingly Eq.(6.10) is identical to Eq.(6.12). This is so because ghost graphs in this gauge result in identical expressions regardless of the orientation of the ghost line.

Goldstone and gauge boson graphs :

**WWV** propagators

$$q^\mu V^{13,14}_{\mu\alpha\beta} + iM_Z S^{7,6}_{\alpha\beta} = 0 \tag{6.13}$$

$$q^\mu V^{15,16}_{\mu\alpha\beta} + iM_Z S^{9,8}_{\alpha\beta} = 0 \tag{6.14}$$

$$q^\mu V^{17,18}_{\mu\alpha\beta} + iM_Z S^{11,10}_{\alpha\beta} + iM_Z S^{13,12}_{\alpha\beta} = gc \left[ \Pi^{9,8}_{\alpha\beta}(1) - \Pi^{9,8}_{\alpha\beta}(2) \right] \tag{6.15}$$

$$q^\mu V^{22}_{\mu\alpha\beta} = 0 \tag{6.16}$$

$$q^\mu V^{45}_{\mu\alpha\beta} = gc \left[ \Pi^{13}_{\alpha\beta}(1) - \Pi^{13}_{\alpha\beta}(2) \right] \tag{6.17}$$

$$q^\mu V^{46}_{\mu\alpha\beta} + iM_Z S^{22}_{\alpha\beta} = 0 \tag{6.18}$$

$$q^\mu V^{47}_{\mu\alpha\beta} + iM_Z S^{23}_{\alpha\beta} = 0 \tag{6.19}$$

**WWH** propagators

$$q^\mu V^{23}_{\mu\alpha\beta} = gc \left[ \Pi^{11}_{\alpha\beta}(1) - \Pi^{11}_{\alpha\beta}(2) \right] \tag{6.20}$$

$$q^\mu V^{24}_{\mu\alpha\beta} + iM_Z S^{14}_{\alpha\beta} = 0 \tag{6.21}$$

$$q^\mu V^{25}_{\mu\alpha\beta} + iM_Z S^{15}_{\alpha\beta} = 0 \tag{6.22}$$

**HZW** and **ZHW** propagators

$$q^\mu V^{29}_{\mu\alpha\beta} + iM_Z S^{17}_{\alpha\beta} = 0 \tag{6.23}$$

and the mirror image graph

$$q^\mu V^{30}_{\mu\alpha\beta} + iM_Z S^{16}_{\alpha\beta} = 0 \tag{6.24}$$

$$q^\mu V^{31}_{\mu\alpha\beta} + iM_Z S^{19}_{\alpha\beta} = 0 \tag{6.25}$$

and the mirror image graph

$$q^\mu V^{32}_{\mu\alpha\beta} + iM_Z S^{18}_{\alpha\beta} = 0 \tag{6.26}$$

$$q^\mu V^{49}_{\mu\alpha\beta} + iM_Z S^{25}_{\alpha\beta} + iM_Z S^{3}_{\alpha\beta} + iM_Z S^{4}_{\alpha\beta} = 0 \tag{6.27}$$

$$q^\mu V^{48}_{\mu\alpha\beta} + iM_Z S^{24}_{\alpha\beta} + iM_Z S^{2}_{\alpha\beta} + iM_Z S^{5}_{\alpha\beta} = 0 \tag{6.28}$$

The rest of the Goldstone boson graphs give :

$$q^\mu V^{19}_{\mu\alpha\beta} + q^\mu V^{20}_{\mu\alpha\beta} + q^\mu V^{21}_{\mu\alpha\beta} = gc \left[ \Pi^{10}_{\alpha\beta}(1) - \Pi^{10}_{\alpha\beta}(2) \right] \tag{6.29}$$

$$-\frac{g^3}{8c} \int Z(k)W(k+p_1)\,(2k+p_1)_\alpha\,(2k+p_1-q)_\beta$$

$$+\frac{g^3}{8c} \int Z(k)W(k-p_2)\,(2k-p_2)_\beta\,(2k-p_2+q)_\alpha$$

$$\tag{6.30}$$

$$q^\mu V^{26}_{\mu\alpha\beta} + q^\mu V^{27}_{\mu\alpha\beta} + q^\mu V^{28}_{\mu\alpha\beta} = gc \left[ \Pi^{12}_{\alpha\beta}(1) - \Pi^{12}_{\alpha\beta}(2) \right] \tag{6.31}$$

$$-\frac{g^3}{8c} \int H(k)W(k+p_1)\,(2k+p_1)_\alpha\,(2k+p_1-q)_\beta$$

$$+\frac{g^3}{8c}\int H(k)W(k-p_2)\,(2k-p_2)_\beta\,(2k-p_2+q)_\alpha \quad (6.32)$$

$$q^\mu V^{33}_{\mu\alpha\beta} + iM_Z S^{21}_{\alpha\beta} = \frac{g^3}{8c}\int H(k)W(k+p_1)\,(2k+p_1)_\alpha\,(2k+p_1-q)_\beta$$
$$-\frac{g^3}{8c}\int Z(k)W(k-p_2)\,(2k-p_2)_\beta\,(2k-p_2+q)_\alpha \quad (6.33)$$

$$q^\mu V^{34}_{\mu\alpha\beta} + iM_Z S^{20}_{\alpha\beta} = \frac{g^3}{8c}\int Z(k)W(k+p_1)\,(2k+p_1)_\alpha\,(2k+p_1-q)_\beta$$
$$-\frac{g^3}{8c}\int H(k)W(k-p_2)\,(2k-p_2)_\beta\,(2k-p_2+q)_\alpha \quad (6.34)$$

We observe that the left-over integrals of the above four equations Eq.(6.30)–Eq.(6.34) cancel.

Adding equations Eq.(6.6)– Eq.(6.34) by parts we arrive at the desired result.

# 7  Conclusions

In this paper we have extended the S–matrix PT with *non–conserved* currents to the case of three boson vertices, with all three incoming momenta *off–shell*. We have outlined in detail how the effective gauge invariant three boson vertices can be constructed, and we have given explicit closed expressions for the vertices $\gamma W^-W^+$, $ZW^-W^+$, and $\chi W^-W^+$ in Eq.(4.33), Eq.(4.34), and Eq.(4.35), respectively. The g.i. three boson vertices were shown to satisfy naive tree–level Ward identities, which relate them to the g.i. gauge boson self–energies previously constructed by the same method in [16]. The derivation of the aforementioned Ward identities relies on the sole requirement of complete gauge invariance of the S-matrix element considered. In particular, no knowledge of the explicit closed form of the three boson vertices involved is necessary, as long as they have been rendered individually $\xi$–independent. The validity of one of these Ward identities has been proved explicitly, through a detailed diagrammatic one-loop analysis, in the context of three different gauges. The above proofs convincingly illustrate the gauge invariant nature of the entire procedure. Most noticeably, the proof of the Ward identity in the unitary gauge supplies additional evidence that the PT endowes the Green's functions computed in the unitary gauge with several desired theoretical properties, as already shown in [29] for the simpler case of the $W$ self-energy.

Of particular interest is the further exploration of the recently advocated connection between the PT and the BFM. Specifically, all cases studied thus far show that the PT Green's functions coincide with the BFM Green's functions, computed at $\xi_Q = 1$. Unfortunately, no general proof of this point exists yet. In section 6.3 we presented a heuristic argument based on the structure of the Feynman rules in this particular gauge, which supports the general validity of this hypothesis, at least at the one-loop level. Due to the lack of a rigorous proof however, additional individual cases may have to be examined. To that end one will have to construct physically relevant Green's functions via the PT and then compare them with the analogous Green's functions of the BFM at $\xi_Q = 1$. The general methodology presented in section 3, and the closed explicit expressions reported in section 4 provide the starting point for such a detailed comparison. Furthermore, the general character of the Ward identities derived in this paper may provide additional clues towards a formal understanding of the PT algorithm. Results in this direction will be reported elsewhere.

# 8 Acknowledgments

One of us (K.P.) would like to thank V. Filippidis for technical support. This work was supported in part by the National Science Foundation Grant No.PHY-9313781.

# 9 Figure Captions

**Figure 1** : One-loop-dressed Feynman graphs for the renormalized $\widehat{\Pi}_{\mu\nu}$ (in a ghost free gauge ) necessary to implement the gauge invariance of the effective potential. All vertices and propagators are fully dressed.

**Figure 2** : The general structure of the part $\widehat{T}(q, p_1, p_2)$ of the S-matrix that depends only on the momentum transfers $q, p_1, p_2$ . The solid lines without orientation represent boson propagators.

**Figure 3** : Graphs contributing pinch parts to the construction of g.i. $Z$ self-energies.

**Figure 4** : The graphs providing pinch parts to the $\gamma WW$ , $ZWW$ and $\chi WW$ vertices in the Feynman gauge. The diagrams pinching the left fermion line as well as all types of crossed diagrams are not shown.

**Figure 5** : The usual graphs contributing to the the $\gamma WW$ and $ZWW$ vertices in the Feynman gauge. Their corresponding expressions are denoted as $V^i_{\mu\alpha\beta}$ in the text. In the unitary gauge only the graphs 1 to 8 , 23 and 29,30 are present; they are denoted as $\mathcal{V}^i_{\mu\alpha\beta}$. In the context of the BFM additional graphs must be included.

**Figure 6** : The usual graphs contributing to the $\chi WW$ vertex in the Feynman gauge. In the text they are denoted as $S^i_{\mu\alpha\beta}$. None of these graphs exist in the unitary gauge. In the BFM additional graphs must be included, whereas graphs containing ghosts are absent.

**Figure 7** : The g.i. self-energies $\widehat{\Pi}^W_{\mu\nu}$, $\widehat{\Pi}^+_\mu$, $\widehat{\Pi}^-_\nu$ and $\widehat{\Pi}^\phi$ (a, b, c, and d, respectively)

**Figure 8** : The vertex , self energy, and monodromic graphs of the S-matrix for the six-fermion process, after the PT rearrangement. Solid lines without orientation represent bosons. All loop expressions are now g.i. and the gauge dependence enters only through the tree propagators of the gauge bosons and their respective scalars. The mirror image and crossed graphs of the monodromic graphs are not shown.

**Figure 9** : Graphs contributing a gauge dependent part of the form $\Delta(q, \xi_Z)$ to the $ZWW$ part of the amplitude.

**Figure 10** : Feynman diagrams contributing to the $W$ self-energy. Graph 13 is particular to the BFM.

**Figure 11** : Box graphs that in the unitary gauge can provide vertex–like pinch parts. In these graphs pinching is triggered through the longitudinal part of the vector bosons' propagators.

**Figure 12** : The additional $\gamma WW$ and $ZWW$ vertex graphs of the Feynman gauge of the background field method. Together with those of **Figure 5** they are denoted by $V^i_{\mu\alpha\beta}$ in section 6.3.

**Figure 13** : The additional $\chi WW$ vertex graphs of the Feynman gauge of the background field method. Together with those of **Figure 5** they are denoted by $S^i_{\mu\alpha\beta}$ in section 6.3. Graphs 2 to 5 replace the ghost graphs 2 to 5 of **Figure 5**, which do not exist in this gauge.

# References

[1] J. M. Cornwall, in Deeper Pathways in High Energy Physics, edited by B. Kursunoglu, A. Perlmutter, and L. Scott (Plenum, New York, 1977), p.683.

[2] J. M. Cornwall, Phys. Rev. D 26 (1982) 1453.

[3] J. M. Cornwall, Phys. Rev. D 38 (1988) 656.

[4] J. M. Cornwall, Physica A 158 (1989) 97.

[5] J. M. Cornwall and J. Papavassiliou , Phys. Rev. D 40 (1989) 3474.

[6] J. Papavassiliou , Phys. Rev. D 47 (1993) 4728.

[7] J. M. Cornwall, R. Jackiw, and E. T. Tomboulis Phys. Rev. D 10 (1974) 2428.
J. M. Cornwall and R. Norton Ann. Phys. (N.Y.) 91 (1975) 106.

[8] Even if one assumes that $\hat{d}$, $\hat{\Gamma}_3$, and $\hat{\Gamma}_4$, are individually g.i., $\Omega$ still displays a residual dependence on the gauge-fixing parameter, stemming from the free part of the gluon propagators. Eq.(1.1) is the necessary condition for the order by order cancellation of this residual gauge dependence

[9] It is only after the gauge invariance of $\Omega$ has been guaranteed that one proceeds to obtain the SD equations for the g.i. Green's functions $\hat{d}$, $\hat{\Gamma}_3$, and $\hat{\Gamma}_4$, by means of a variational principle. To that end, one extremizes independently the variations of $\Omega(\hat{d}, \hat{\Gamma}_3, \hat{\Gamma}_4)$ with respect to $\hat{d}$, $\hat{\Gamma}_3$, and $\hat{\Gamma}_4$, e.g. $\frac{\delta\Omega}{\delta\hat{\Delta}} = 0$, $\frac{\delta\Omega}{\delta\hat{\Gamma}_3} = 0$, and $\frac{\delta\Omega}{\delta\hat{\Gamma}_4} = 0$, imposing Eq.(1.4) as an additional constraint.

[10] It is important to emphasize that the Ward identities of Eq.(1.4) make no reference to ghosts; they are valid however also in the context of covariant gauges after the application of the PT.

[11] R. Jackiw and K. Johnson Phys. Rev. D 8 (1973) 2386.
J. M. Cornwall and R. Norton Phys. Rev. D 8 (1973) 3338.
E. Eichten and F. Feinberg Phys. Rev. D 10 (1974) 2428.

[12] J. Papavassiliou, Phys. Rev. D 41 (1990) 3179.

[13] G. Degrassi and A. Sirlin, Nucl. Phys. B 383 (1992) 73.
G. Degrassi and A. Sirlin, Phys. Rev. D 46 (1992) 3104.

[14] G. Degrassi, B. Kniehl, and A. Sirlin, Phys. Rev. D 48 (1993) R3963.

[15] J. Papavassiliou and C. Parrinello, Phys. Rev. D 50 (1994) 3059.

[16] J. Papavassiliou, Phys. Rev. D 50 (1994) 5958.

[17] K.Hagiwara, S.Matsumoto, C.S.Kim, Proceedings of the 14'th International Workshop, *Weak Interactions and Neutrinos*, Edited by J.E.Kim and S.K.Kim, (World Scientific, 1994), p.19.
K.Hagiwara, D.Haidt, C.S.Kim and S.Matsumoto, Z. Phys. C64 (1994) 559.

[18] K. J. F. Gaemers and G. J. Gounaris, Z. Phys. C 1 (1979) 259.
K. Hagiwara, R.D. Peccei, D. Zeppenfeld, K. Hikasa, Nucl. Phys. B 282 (1987) 253.

[19] U. Baur and D. Zeppenfeld, Nucl. Phys. B 308 (1988) 127.
U. Baur and D. Zeppenfeld, Nucl. Phys. B 325 (1989) 253.
E. N. Argyres, F.K. Diakonos, O. Korakianitis, C.G. Papadopoulos, W.J. Stirling, Phys. Lett. B 272 (1991) 431.
E.N. Argyres, G. Katsilieris, O. Korakianitis, C.G. Papadopoulos, K. Philippides,

W.J. Stirling, Phys. Lett. B 280 (1992) 324.

F.K. Diakonos, O. Korakianitis, C.G. Papadopoulos, K. Philippides, W.J. Stirling, Phys. Lett. B 303 (1993) 177.

[20] S. J. Brodsky and J. R. Hiller, Phys. Rev. D 46 (1992) 2141.

M. Clauson, E. Fahri, and R. L. Jaffe Phys. Rev. D 34 (1986) 873.

G. Belanger, F. Boudjema, D. London, Phys. Rev. Lett. 65 (1990) 2943.

F. Boudjema, Phys.Rev. D 36 (1987) 969.

J. Wudka, Proceedings of the Workshop *Search for New Phenomena at Colliding Beam Facilities*, Oct 2-3, 1992, editor: Jeanne Rogers.

J. Wudka, Proceedings of the XXVIIIth Rencontre de Moriond, *Electroweak Interactions and Unified Theories*, Les Arcs, Savoie, France, March 13-20, 1993, editor: J. Tran Thanh Van; pp. 441–448. Editions Frontieres, Gif-sur-Yvette, France (1993).

[21] K. Fujikawa, B. W. Lee, and A. I. Sanda, Phys. Rev. D 6 (1972) 2923.

[22] E.N. Argyres, G. Katsilieris, A.B. Lahanas, C.G. Papadopoulos, and V.C. Spanos, Nucl. Phys. B 391 (1993) 23.

[23] J. Papavassiliou and K. Philippides, Phys. Rev. D 48 (1993) 4255.

[24] L. F. Abbott, Nucl. Phys. B 185 (1981) 189, and references therein.

[25] A. Denner, S. Dittmaier, and G. Weiglein, Phys. Lett. B 333 (1994) 420.

S. Hashimoto, J. Kodaira, Y. Yasui, and K. Sasaki, Phys. Rev D 50 (1994) 7066.

E. de Rafael and N. J. Watson, (unpublished).

[26] Of course, one can apply the PT in the context of the BFM, to eliminate the residual $\xi_Q$-dependence, when $\xi_Q \neq 1$. In that case one arrives again at precisely the same unique results as in any other gauge-fixing scheme [27].

[27] J. Papavassiliou, Phys. Rev. D 51 (1995) 856.

[28] N. Jay Watson, *Universality of the Pinch Technique Gauge Boson Self Energies*, CPT–94–P–3133, hep-ph 9412319.

[29] J. Papavassiliou and A. Sirlin, Phys. Rev.; D 50,5951 (1994).

[30] K. Aoki, Z. Hioki, R. Kawabe, M. Konuma, T. Muta, Prog. Theor. Phys. Suppl. 73 (1982) 1.

[31] We emphasize again that, since the quantities constructed via the PT are $\xi$-independent, any choice for the parameters $\xi$ is legitimate.

[32] The diagrams where the pinching occurs at the fermion line on the left are exactly analogous to those where the pinching occurs at the fermion line on the right, and are not shown.

[33] We remind the reader that the tadpole diagrams must also be included in the definition of the g.i. self-energies.

[34] (ii) and (iii) are one-particle reducible graphs. The graphs in category (ii) become disconnected only after a gauge-boson line is cut. The graphs in (iii) become disconnected in an additional way, namely by cutting the off-shell fermionic propagator.

[35] For the graphs with scalars attached to the external fermion lines the above factor can be pulled out after we use the relations of Eq.(2.17).

$$\hat{\Pi}_{\mu\nu} = \tfrac{1}{2} \quad \text{[diagram]} \quad + \quad \tfrac{1}{2} \quad \text{[diagram]}$$

Figure 1

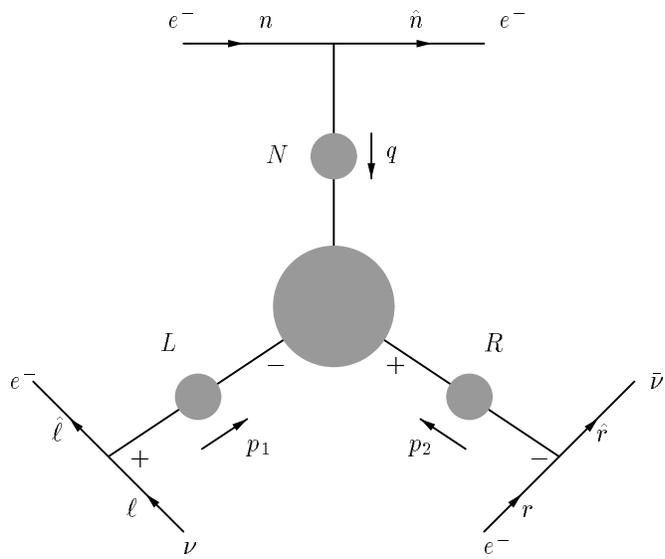

Figure 2

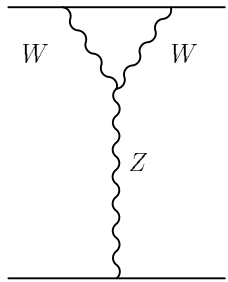
(a)

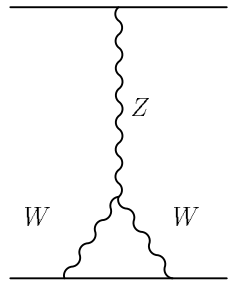
(b)

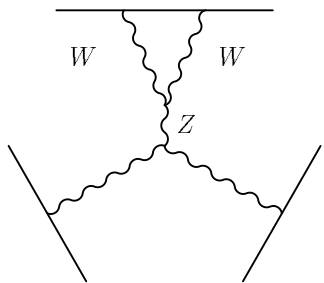
(c)

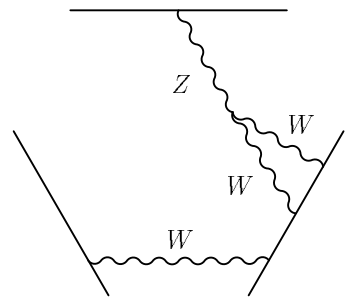
(d)

Figure 3

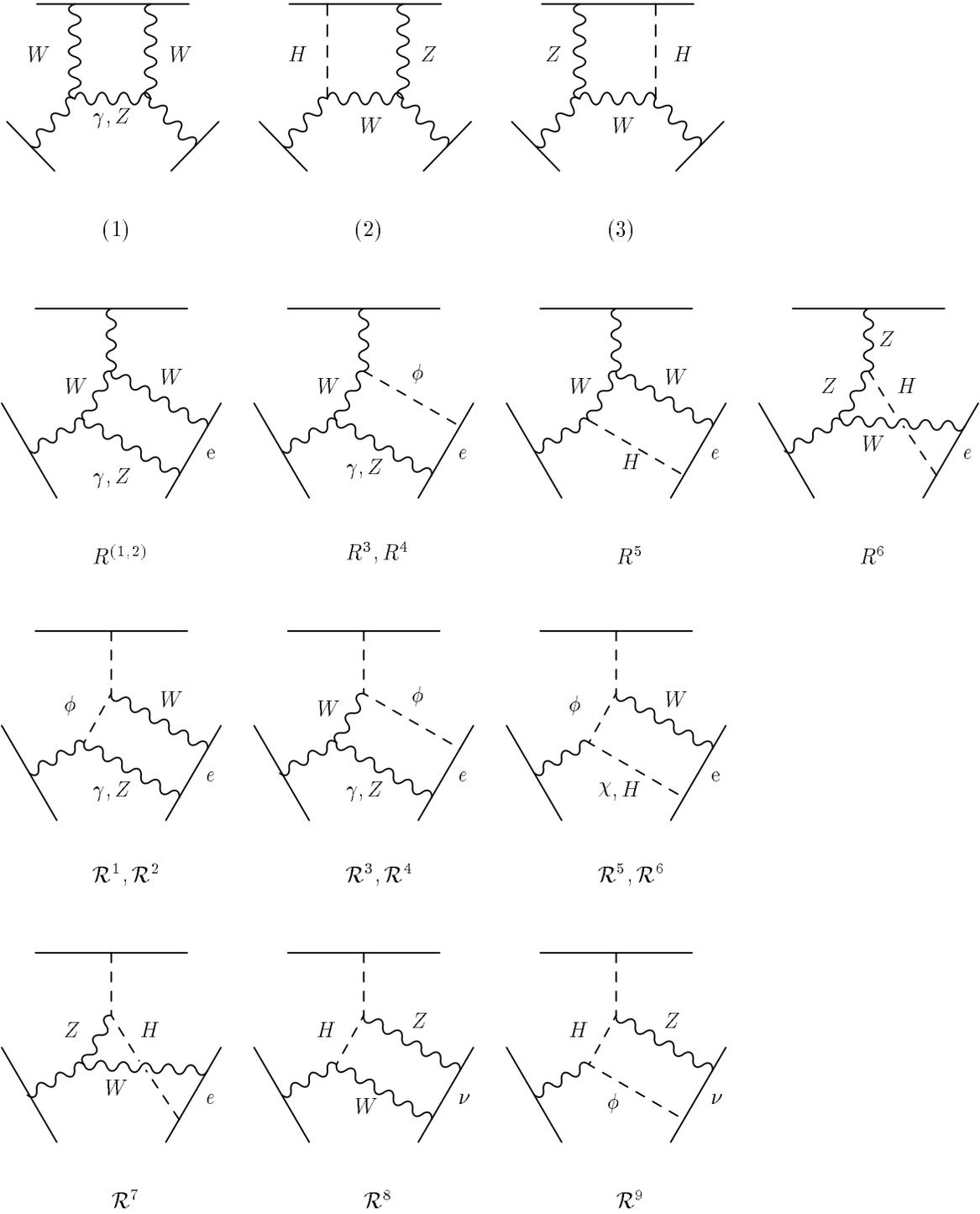

Figure 4

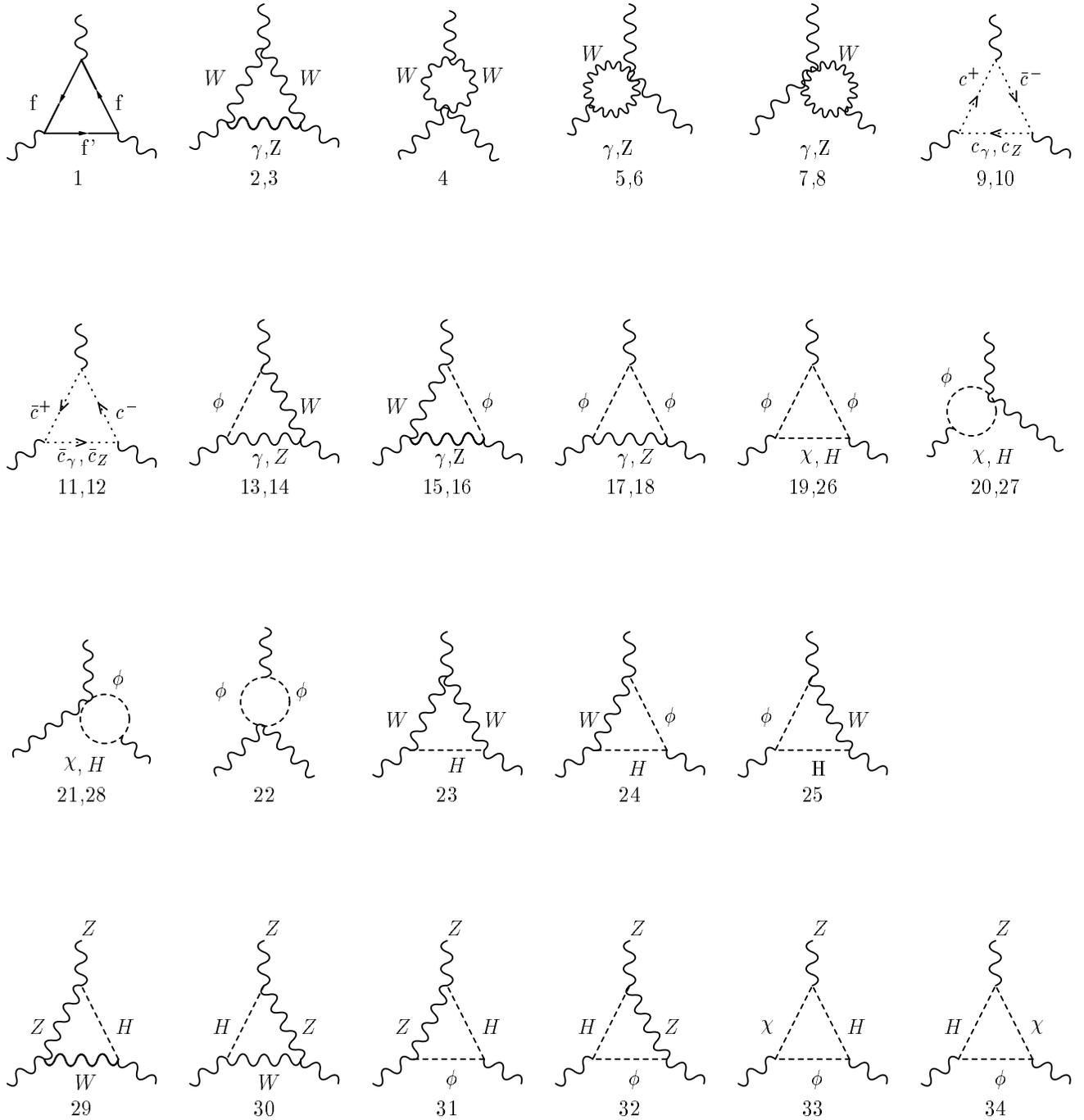

Figure 5

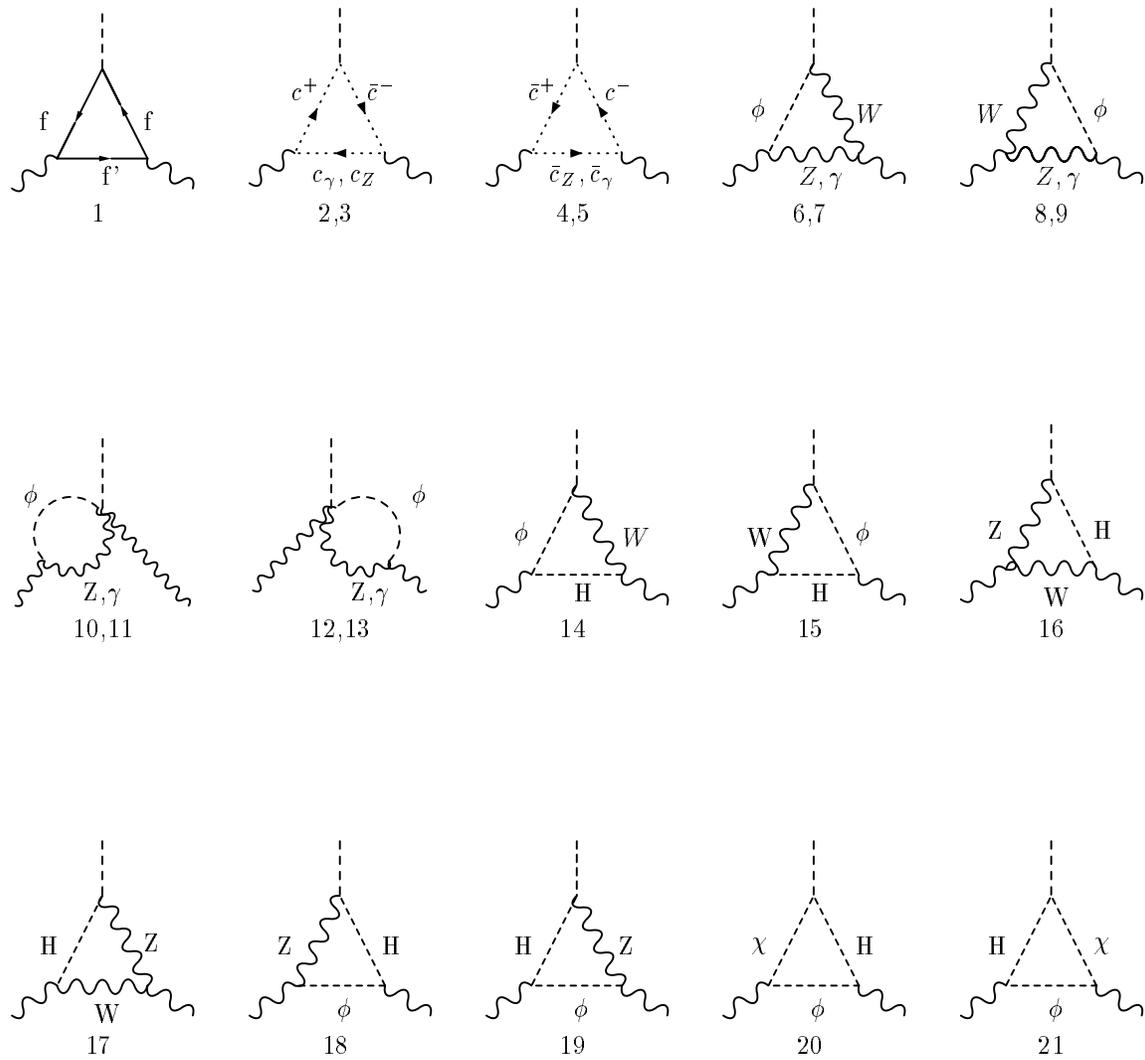

Figure 6

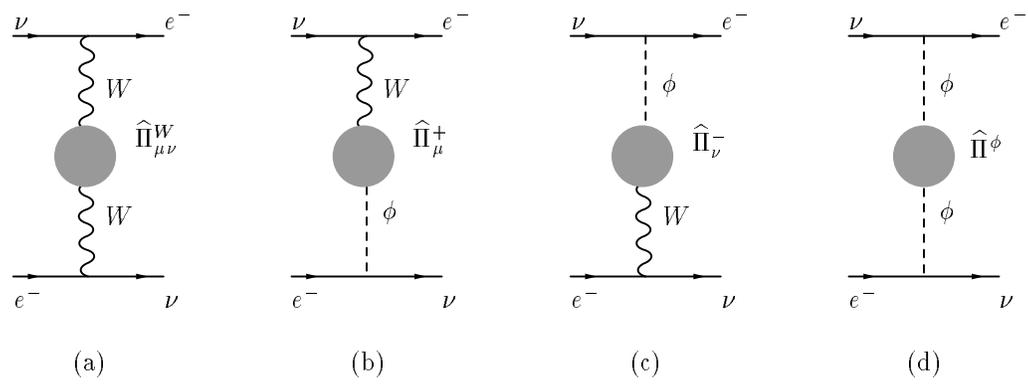

Figure 7

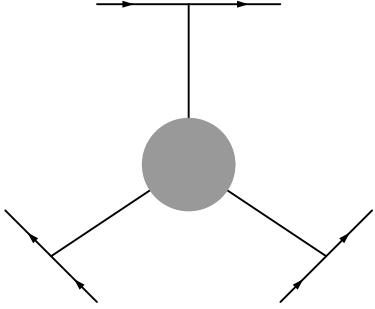

(a)

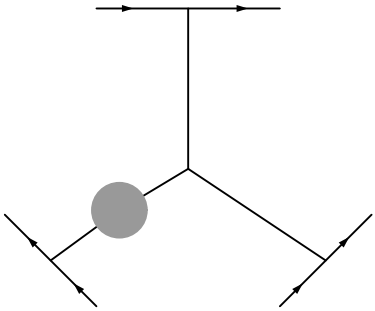 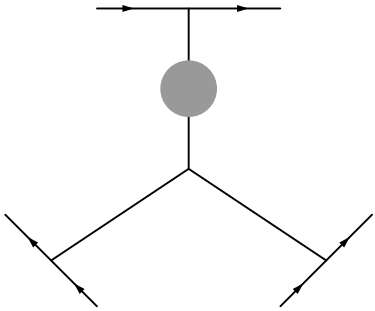 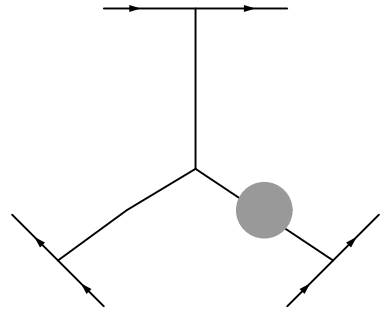

(b) (c) (d)

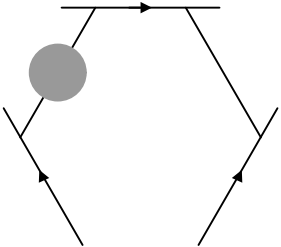 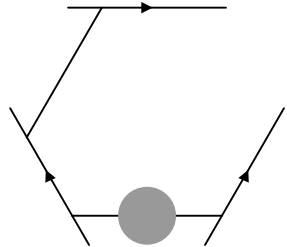 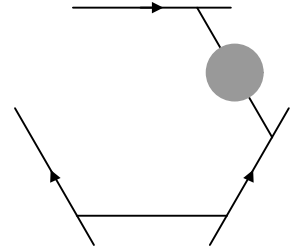

(e) (f) (g)

Figure 8

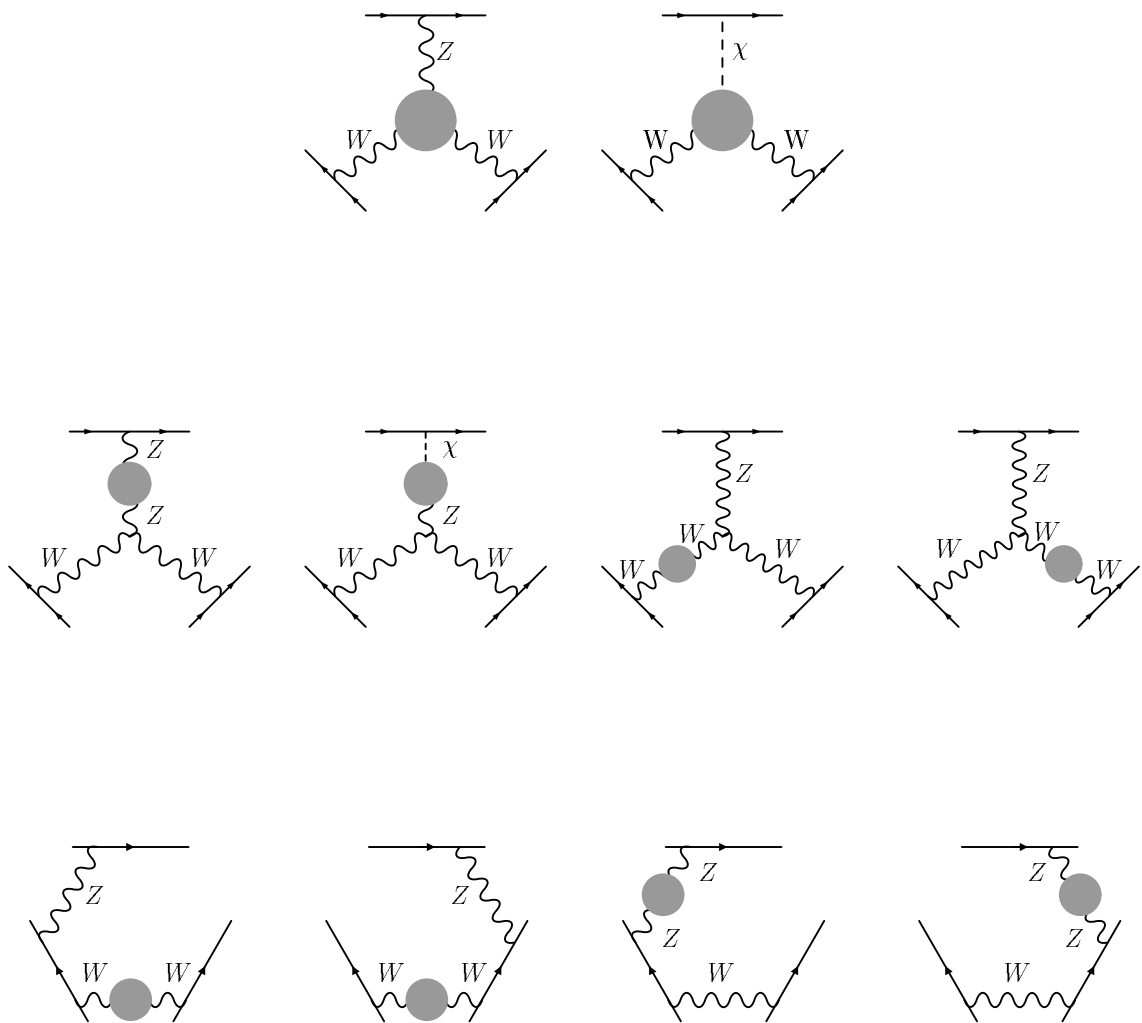

Figure 9

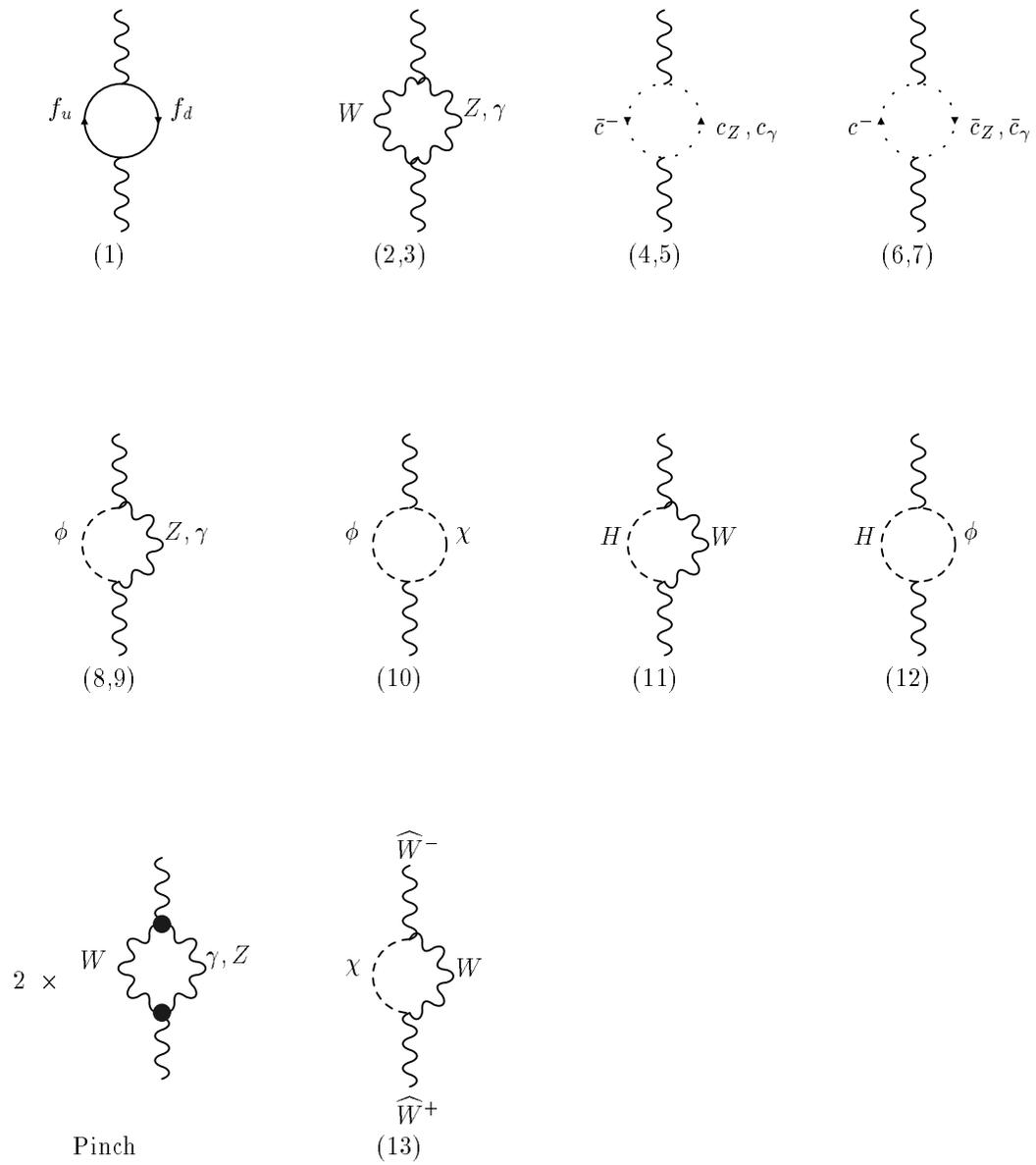

Figure 10

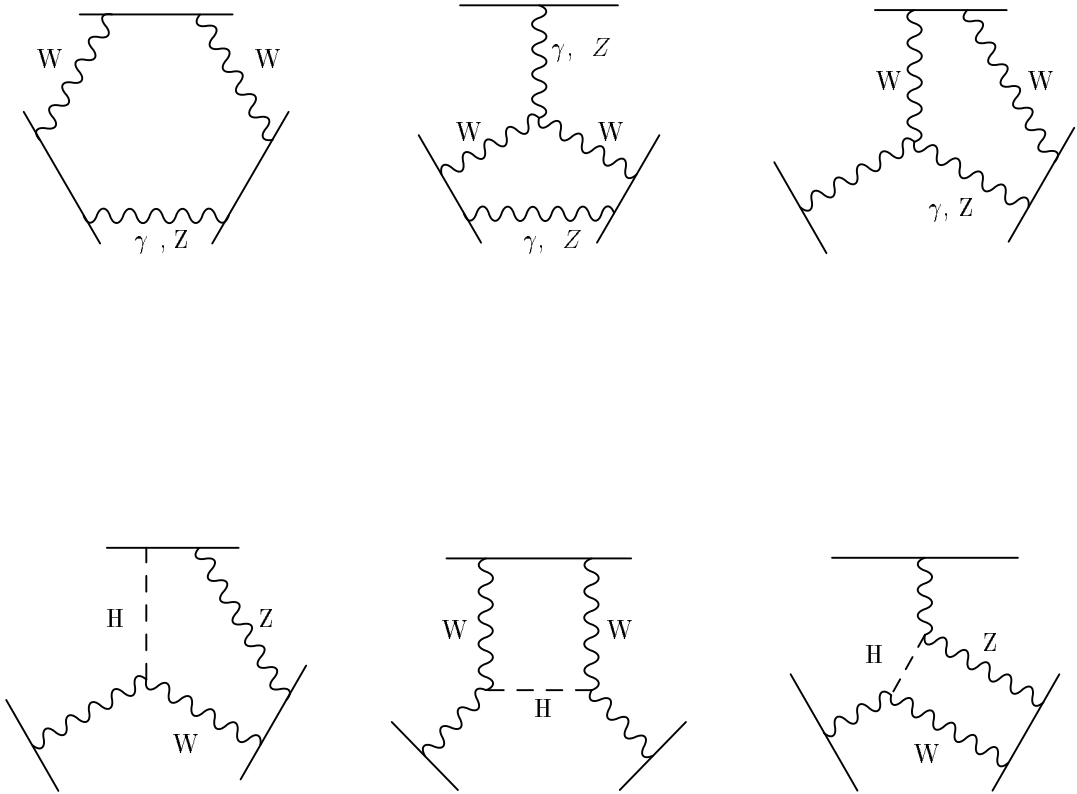

Figure 11

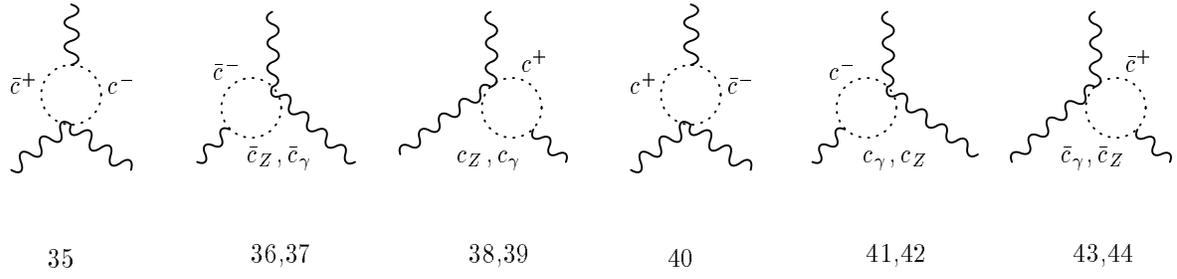

35        36,37        38,39        40        41,42        43,44

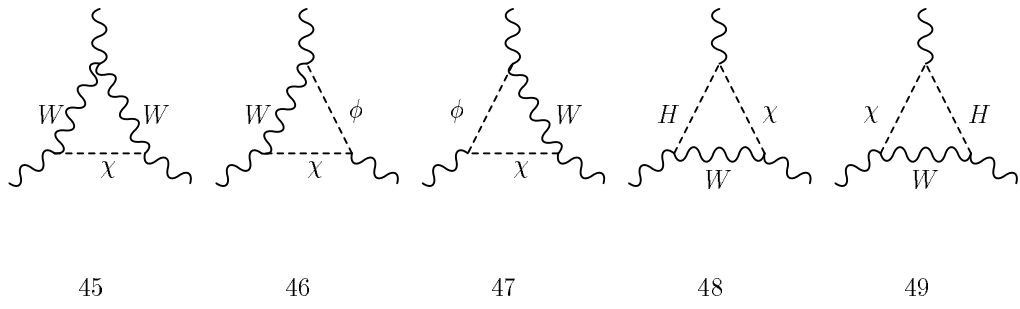

45        46        47        48        49

Figure 12

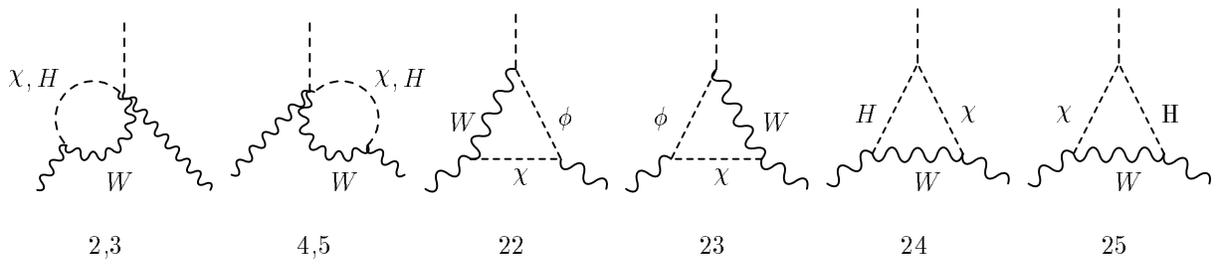

Figure 13